\documentclass[12pt]{aastex62}
\usepackage{xcolor}
\usepackage{soul}
\usepackage{amsmath}
\usepackage{verbatimbox}
\usepackage{subfloat}
\usepackage{listings}

\begin{document}

\title{Toward the Detection of Relativistic Image Doubling in Water Cherenkov Detectors}
\author[0000-0003-4786-2348]{Neerav Kaushal}
\email{kaushal@mtu.edu}
\affiliation{Michigan Technological University\\1400 Townsend Drive\\Houghton, MI 49931,USA}
\author[0000-0002-4505-6599]{Robert J.~Nemiroff}
\email{nemiroff@mtu.edu}
\affiliation{Michigan Technological University\\1400 Townsend Drive\\Houghton, MI 49931,USA}

\begin{abstract}

When a gamma or cosmic ray strikes the top of Earth's atmosphere, a shower of secondary particles moves toward the surface. Some of these secondary particles are charged muons that subsequently enter Water Cherenkov Detectors (WCDs) on the ground. Many of these muons, traveling near the speed of light in vacuum, are moving faster than the speed of light in water and so trigger isotropic Cherenkov radiation in the WCDs. Inside many WCDs are photomultiplier tubes (PMTs) that detect this Cherenkov radiation. When the radial component of the speed of a muon toward a PMT drops from superluminal to subluminal, the PMT will record Cherenkov light from a little-known optical phenomenon called Relativistic Image Doubling (RID). Were the RID-detecting PMTs replaced by high resolution video recorders, they would see two Cherenkov images of the muon suddenly appear inside the tank, with one image moving with a velocity component toward the recorders, the other away. Even without a video, the RID phenomenon will cause different PMTs to record markedly different light curves for the same muon. In this paper, we present a study hoping to inspire the explicit detection and reporting of RID effects in WCDs. We consider three example cases of muon RIDs in High-Altitude Water Cherenkov (HAWC)-like systems: vertical, horizontal and oblique. Monte Carlo simulations show that RID effects in HAWC-like systems are not rare -- they occur for over 85 percent of all muon tracks. 
\end{abstract}

\section{Introduction} \label{sec:intro}

Air showers of fundamental particles result when a single high energy particle such as a gamma-ray strikes the top of the Earth's atmosphere \citep{1933ZPhy...82..151R}. A common type of secondary shower particle is the charged muon which decays in 2.2 microseconds in its own inertial frame \citep{2018PhRvD..98c0001T}, but in the frame of the Earth, its relativistic speed can cause it to last much longer. Long time delays allow relativistic muons to reach the surface before decaying.

At the surface, secondary muons may enter vats of water designed to detect them. The key to detection is the Cherenkov radiation \citep{1937PhRv...52..378C} that they cause in the water tanks, which is created because the speed of the charged muon is much greater than the speed of light in water. 

Objects moving faster than light in a given medium may create images in non-classical ways. One recently studied example is Relativistic Image Doubling (RID) \citep{1971Sci...173..525C, 2015PASA...32....1N, 2018AnP...53000333N, 2019AAS...23325101N}. With RID, objects moving superluminally in a medium (faster than the medium speed of light) can appear twice simultaneously to an observer. The RID illusion starts when the moving object's speed toward this observer drops from superluminal to subluminal. At that time, two bright images of the objects suddenly appear and diverge. RID effects have recently been found in the lab \citep{2016SciA....2E1691C} and have been hypothesized to help explain light curves in gamma-ray bursts \citep{2019ApJ...883...70H}. Most recently, it has been suggested that RID effects occur and can be commonly found in images of air showers by Imaging Atmospheric Cherenkov Telescopes \citep{2020ApJ...889..122N}. 

In this work, it will be shown that RID effects not only exist but are common for muon tracks in ground-based Water Cherenkov Detectors (WCDs). WCDs exist in many cosmic gamma-ray detectors today, including those deployed by Auger \citep{2015ApJ...802..111A}, HAWC \citep{2015EPJWC.10501003B}, Kamiokande \citep{ 1984AIPC..114...54A}, and IceCube \citep{2017JInst..12P3012A}. 

The paper is structured as follows. In Section 2 the conceptual basis for RIDs are reviewed in relation to how they might affect WCDs. Also we develop the mathematical framework behind the concepts, assuming the muon takes an arbitrary track in the WCD, and the detector is located at an arbitrary position on the WCD floor. In Section 3 we calculate what results would be expected from three different types of detectors: a video detector that records both the brightness of the muon track and its angular position with time, a digital camera that records the brightness as a function of angular position but not time, and a PMT that records brightness as a function of time but not angular position. Also, we show simulation results for an example WCD tank that is similar to the tanks in the High-Altitude Water Cherenkov (HAWC) observatory. In Section 4 it is shown how prevalent RIDs are in WCDs like HAWC, and how RID resolution might help decode the muon track. In Section 5 we discuss various aspects of RID in regard to its applications and future developments. Finally some of the codes used for the simulations are shown in Appendix.

\section{RID: Concepts and Mathematics}

Following  \citet{2020ApJ...889..122N}, several RID concepts are found to be relevant here. Consider a cosmic ray muon travelling with a speed $v > c_w$, where $c_w$ is the speed of light in water, and entering the top of a Water Cherenkov Detector (WCD) tank filled up to height $H$ and leaving the tank through the bottom. The speed $v$ of the muon is considered here to be constant during its entire track across the WCD. The muon enters the WCD at time $t=0$ and causes the emission of isotropic Cherenkov light in its immediate wake. Although not a direct image of the muon, this light will be referred to as a ``Cherenkov image".

\begin{figure}[h]
    \centering
    \hspace*{-1.0cm}
    \includegraphics[scale=0.6]{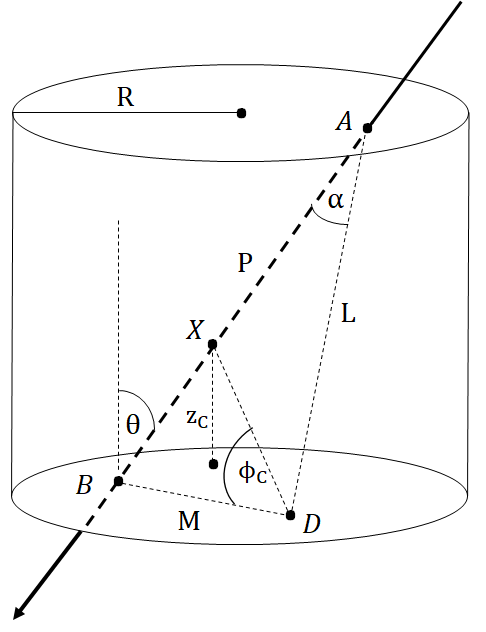}
    \caption{A muon entering the top of the WCD tank through $A$ and leaving the bottom through $B$. The path length of the muon inside the tank is given by $P$ while the distances of the detector $D$ from $A$ and $B$ are given by $L$ and $M$ respectively. $\alpha$ is the angle between the muon track and the detector through the entry point $A$.}
    \label{diagram_1}
    \end{figure}

The light is assumed to be seen here by a video detector that tracks both angular position and brightness in time. The detector, on the floor of the WCD, is assumed to be located at a distance $L$ from the point where the muon enters the WCD and at distance $M$ from the point where it exits the WCD. The path length traveled by the muon inside the tank is given by $P$. The height of the muon from the tank bottom at any time $t$ during its course in the tank is given by $z$. This is shown in Figure~\ref{diagram_1}.

Because of the superluminal speed (meaning $v > c_w$) of the muon, some muon tracks will appear to have non-classical attributes. Specifically, some muon tracks will be seen by the video detector to undergo relativistic image doubling (RID).

The velocity of the muon at any point in its trajectory can be considered to have two components relative to each detector -- one radially toward (or away) from the detector ($v_r$), and one perpendicular to the line connecting the detector to the muon. If the detector is nearly in the path of the incoming muon, then $v_r > c$. Conversely, when the muon is closest to the detector, $v_r = 0$. For this incoming muon, there will be a location in its track where $v_r$ drops from superluminal to subluminal at which the detector will see an RID. This is equivalent to that detector passing from outside to inside the Cherenkov cone of the muon. The video will record the muon's Cherenkov images suddenly appearing at the location of the RID, typically away from the top or any edge of the WCD. The muon, as shown by its Cherenkov images, will then appear on the video to be at two places simultaneously, with one image moving along the original muon track, while a second image moving backwards along the entry track of the muon.

For some detector locations, the muon's radial speed toward the detector is always subluminal, even though the total speed of the muon is always superluminal. These cases are the most classical, as the detector observes the muon move from the entry point in the tank, downward. In this case, no RID would be recorded by the video detector.

Let's now concentrate on detectors which do see RID events. From the point of view of these detectors, muons will first traverse a region starting from their point of entry into the WCD and extending down to $z_C$ where $v_r = c_w$. In this region, the speed of the muon toward the detector $(v_r)$ is {\it faster} than the speed of the Cherenkov radiation it causes. When eventually seen by the detector, this part of the muon track will be seen time-backwards, meaning that Cherenkov radiation emitted increasingly earlier along the track will reach the detector at increasingly later times. This part of the track will appear to go {\it up} from $z_C$. After the muon has descended down past $z_C$ however, the radial component of its speed will be {\it slower} than its Cherenkov light and so this part of the muon track will appear normally: time forwards and headed down. Thus the muon is first seen by the detector at height $z_C$ and {\it not} at the point of its entry in the tank.

The total time from when the muon enters the tank to when the detector records Cherenkov radiation from the muon's trail can be broken up into two times, the time $t_{muon}$ taken by the muon to descend to a height $z$, and the time $t_{radiation}$ taken by the Cherenkov light to travel from $z$ to the detector. This combined time is denoted by $t_{total}$.

The critical height $z_C$ where the muon is first seen by the detector occurs at time $t_{min}$, when $t_{total}$ is a minimum. This can be found by solving $dt_{total}/dz=0$ for $z$. For a muon entering the tank from the top and leaving through the bottom, $z_C$ is given by
    \begin{equation} \label{zc}
    {
    z_C = H - \Big(L \cos \alpha - \frac{c_w L \sin \alpha}{\sqrt{v^2-c_w^2}} \Big) \cos \theta ,
    }
    \end{equation}
where $\theta$ is the angle the muon track makes with the vertical and $\alpha$ is the angle between the detector and the muon track through point $A$.
The angular location, $\phi_C$, where the muon is first seen by the detector, is the angle between the line joining the detector $D$ with the location $X$ of critical height $z_C$ on the muon path (i.e. $DX$) and the line joining the detector with the exit point of muon at the WCD floor (i.e. $BD$). The distance between the detector $D$ and the location of first RID on the muon track $X$ is given by

    \begin{equation} \label{temp}
    \begin{split}
        DX & = \sqrt{AD^2+AX^2-2 AX \cdot AD \cos \alpha} \\
           & = \sqrt{L^2 + \Big(P-\frac{z_C}{\cos \theta}\Big)^2 - 2 L \cos \alpha \Big(P-\frac{z_c}{\cos \theta}\Big)}.
    \end{split}
    \end{equation}
    
Therefore,
 
    \begin{equation} \label{phic}
    \begin{split}
        \phi_C & = \arccos{\Big(\frac{BD^2 + DX^2 - BX^2}{2 \cdot BD \cdot DX}\Big)} \\
               & = \arccos{\Bigg(\frac{ \left(P^2+L^2+M^2\right)\cos \theta +2 L \cos \alpha  (z_C-P \cos \theta )-2 P z_C}{2 M \cos          \theta  \sqrt{L^2-2 L \cos \alpha  \left(P-\frac{z_C}{\cos \theta }\right)+\left(P-\frac{z_C}{\cos          \theta }\right)^2}}\Bigg)}.
    \end{split}
    \end{equation}
Note that an angle $\phi_C=0^{\circ}$ corresponds to the line joining the detector and the exit point $B$. This is shown by the line segment $BD$ in Figure~\ref{diagram_1}.

{\it Near} $z_C$, the muon's speed toward the detector almost equals the speed of its liberated Cherenkov light. Therefore a relatively large amount of this light arrives at the detector ``bunched up" -- over a very short period of time  \citep{2020ApJ...889..122N}. It can then be said that at height $z_C$, the detector is on the ``Cherenkov Cone". Note that since Cherenkov light is emitted uniformly and isotropically along the muon's trajectory in the frame of water, the Cherenkov Cone is a detector-dependent phenomenon.

After $t_{min}$, two images of the muon's Cherenkov light are observed simultaneously by the detector at heights $z_{\pm}$ from the bottom of the tank, where $z_+$ and $z_-$ are always above and below $z_C$ respectively,
    \begin{equation} \label{zpm}
    {
    z_{\pm} = H - \Bigg(\frac{c_w^2 t_{total} v-L v^2 \cos \alpha \pm \sqrt{v^2 \left(c_w^2 \left(L^2+t_{total}^2 v^2\right)-2 c_w^2 L t_{total} v \cos \alpha +L^2 v^2 \cos ^2\alpha-L^2 v^2\right)}}{c_w^2-v^2} \Bigg)\cos \theta.
    }
    \end{equation}
The angular locations of the two images as seen from the detector are given by
    \begin{equation} \label{phipm}
    {
    \phi_{\pm} = \arccos{\Bigg(\frac{ \left(P^2+L^2+M^2\right)\cos \theta +2 L \cos \alpha  (z_{\pm}-P \cos \theta )-2 P z_{\pm}}{2 M \cos          \theta  \sqrt{L^2-2 L \cos \alpha  \left(P-\frac{z_{\pm}}{\cos \theta }\right)+\left(P-\frac{z_{\pm}}{\cos          \theta }\right)^2}}\Bigg)}.
    }
    \end{equation}

Once the height and time of each image of the pair is known, their velocities can be calculated by taking $v_{\pm}=d z_{\pm} / dt_{total}$. The apparent angular speeds ${\omega}_{\pm}$ of these images can be computed by dividing the transverse components of their speeds by their distances from the detector. The apparent brightness of an image is computed by integrating the instantaneous brightness of the track visible to the detector over a uniform time interval. However, since relatively long path lengths are seen over a uniform time interval when $v_r \sim c_w$, the apparent brightness $b$ of each image is proportional to its angular speed. At the extreme, when $v_r = c_w$, the apparent angular speed diverges and the length of the path visible to the detector will formally diverge even over an arbitrarily small time interval. In a separate effect, an image brightness falls by the square of its distance to the detector. Together, these effects yield image brightness of
    \begin{equation} \label{brightness}
    {
    b_{\pm}  \propto { { {\omega_{\pm}}} \over { L^2 + z_{\pm}^2 } }
    }
    \end{equation}
Although it is cumbersome to expand this equation analytically because of the complexity of  Eq. \ref{zpm}, it is straightforward to use this equation numerically. Formally, the brightness of the shower as recorded by the detector mathematically diverges, when the shower appears at ``criticality" when $v_r = c_{w}$ at $z_{\pm} = z_C$. This divergence is mitigated in practice, however, by the muon's track being of finite angular size.

\section{Cases} \label{sec:RID}

\begin{figure}[h]
    \centering
    \hspace*{-1.0cm}
    \includegraphics[scale=0.4]{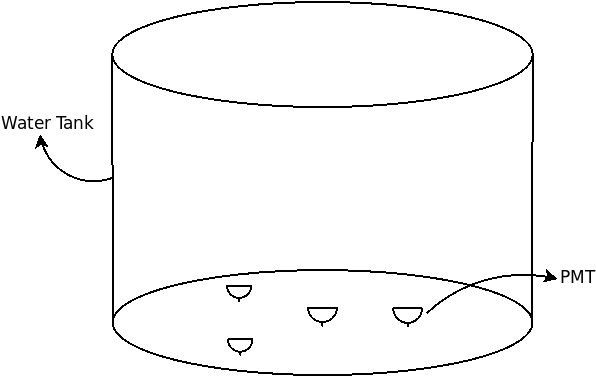}
    \caption{A Three-Dimensional representation of a HAWC-like Water Cherenkov Detection (WCD) tank. The placement of the PMTs on the tank floor is also shown. A bird's eye view of this tank is shown in Figure~\ref{cross_section} along with a track for a muon entering horizontally.}
    \label{figure_hawc_system}
    \end{figure}

Three example cases are examined here. The WCD tank is assumed to be similar to that used by HAWC. This tank is  a cylinder with radius $R = 3.65$ meters and height $5$ metres filled with water up to height $H = 4.5$ meters \citep{2013ICRC...33.1155S}. Four PMTs are assumed to be on the floor of the tank with one in the center and three uniformly spread in a Y-pattern, each at a distance of 1.85 meters from the central PMT (H\"{u}entemeyer, private communication). A three-dimensional view of the system is depicted in Figure~\ref{figure_hawc_system}. Inside this WCD, three types of detectors will be considered. The first is a video detector with the ability to resolve Cherenkov images in both time and angle. The second is a static camera with angular but not time resolving abilities and the third is a PMT which can resolve brightness with time but not with angle.

\subsection{Vertical Incidence}

\begin{figure}[h]
            \centering
            \hspace*{-1.0cm}
            \includegraphics[scale=0.4]{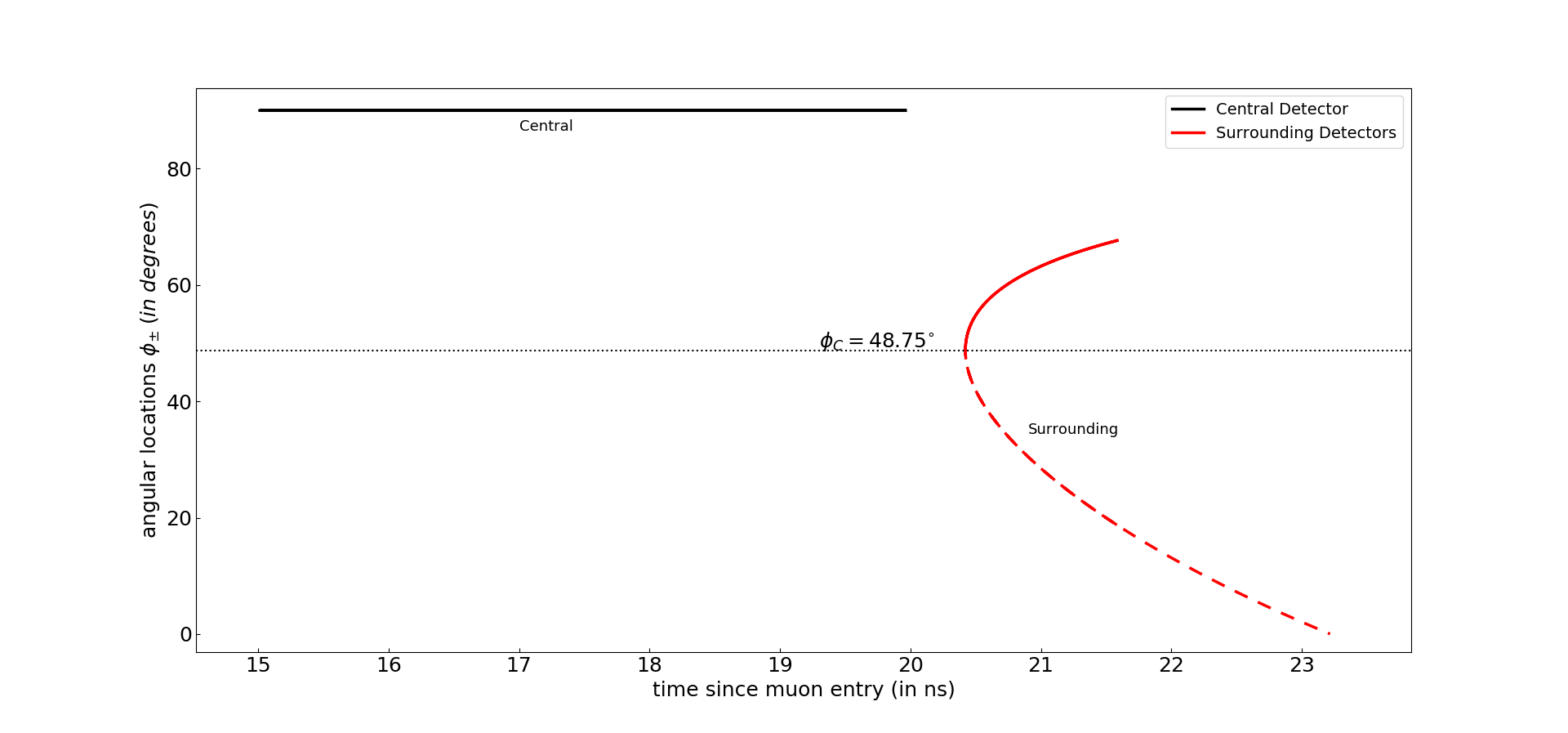}
        \caption{A plot of angular location of Cherenkov images versus time for a muon entering the WCD straight down and striking the central detector at the bottom of the tank. A solid curve represents the Cherenkov image moving upwards while a dashed curve means that the image is moving downwards. The three surrounding detectors are at 1.85 meters from the central one and thus the red curve shown is the same for all three detectors. $\phi_C$ denotes the angular location for the first RID and corresponds to the critical height $z_C$.}
            \label{vertical_phi_vs_t}
        \end{figure}

The first case considered is particularly simple and formulated to calibrate intuition. In this case, the muon enters the WCD vertically moving straight down and directly striking the central detector. 
The detectors are all considered to be video detectors, at first. The only Cherenkov image visible to the central detector in this scenario is the one moving upwards, which, counter-intuitively, is in the opposite direction of the actual downward motion of the muon! The three surrounding detectors will see the same Cherenkov image motions as they are all at the same distance from the muon track as well as the central detector, and symmetrically oriented. The first event these detectors will see is the pair creation event or RID. After that, each detector will record the two Cherenkov images of the muon track moving in geometrically opposite directions along the original track. The video detector in this case will see the development of Cherenkov images along the muon track with time. Data extracted from the resulting video would show the angular position of the Cherenkov images versus time as shown in Figure~\ref{vertical_phi_vs_t}.

\begin{figure}[t]
            \centering
            \hspace*{-1.0cm}
            \includegraphics[scale=0.4]{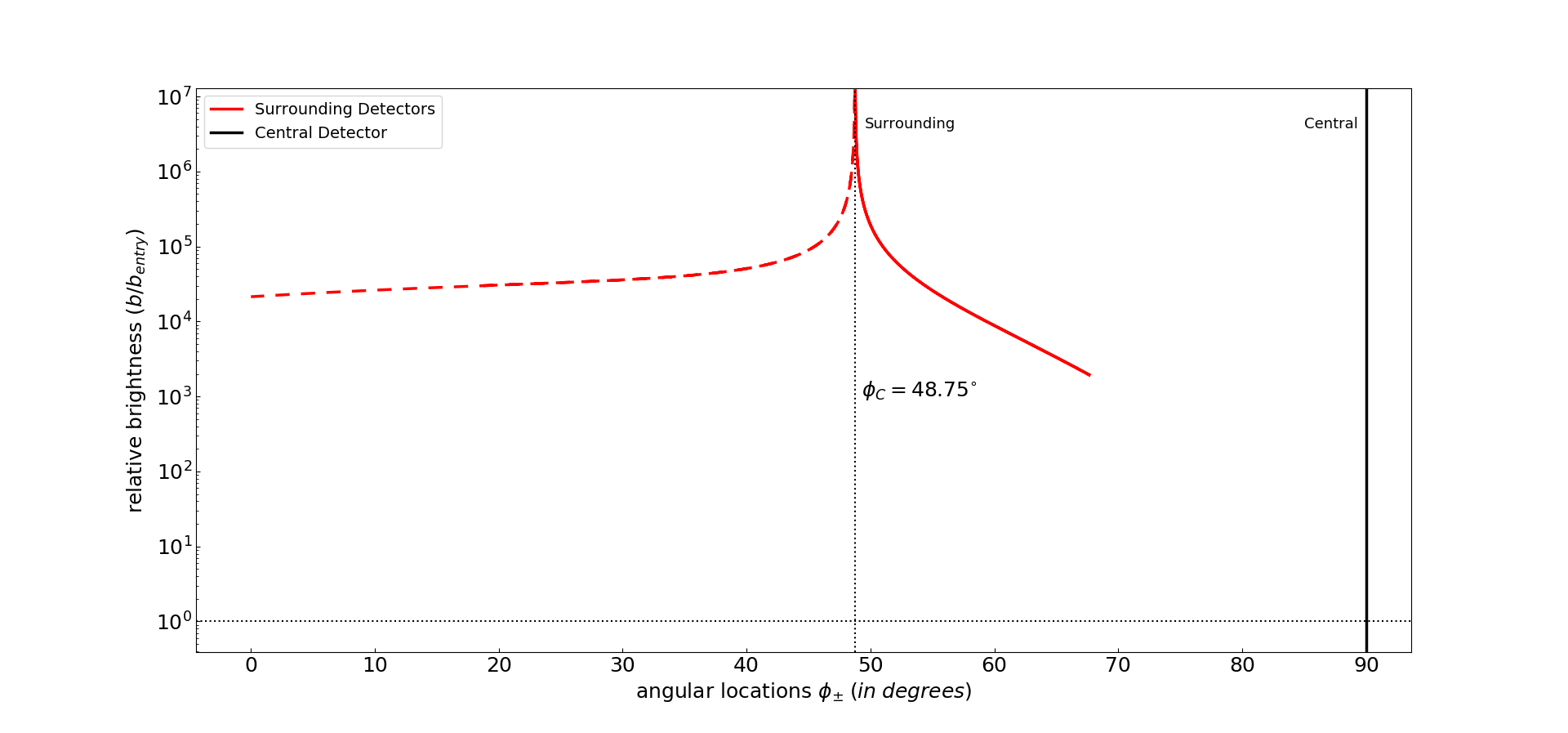}
        \caption{A graph of relative image brightness versus the angular locations of the Cherenkov images of a muon moving straight down and striking the central detector. The central detector always sees a Cherenkov image directly above it at $\phi=90$ degrees.}
            \label{vertical_b_vs_phi}
        \end{figure}

Simpler than a video, a static digital camera image will record a line that encodes the variation of brightness of Cherenkov images with their corresponding angular locations. This can be seen from the inspection of Figure~\ref{vertical_b_vs_phi}.
        
\begin{figure}[h]
            \centering
            \hspace*{-1.0cm}
            \includegraphics[scale=0.4]{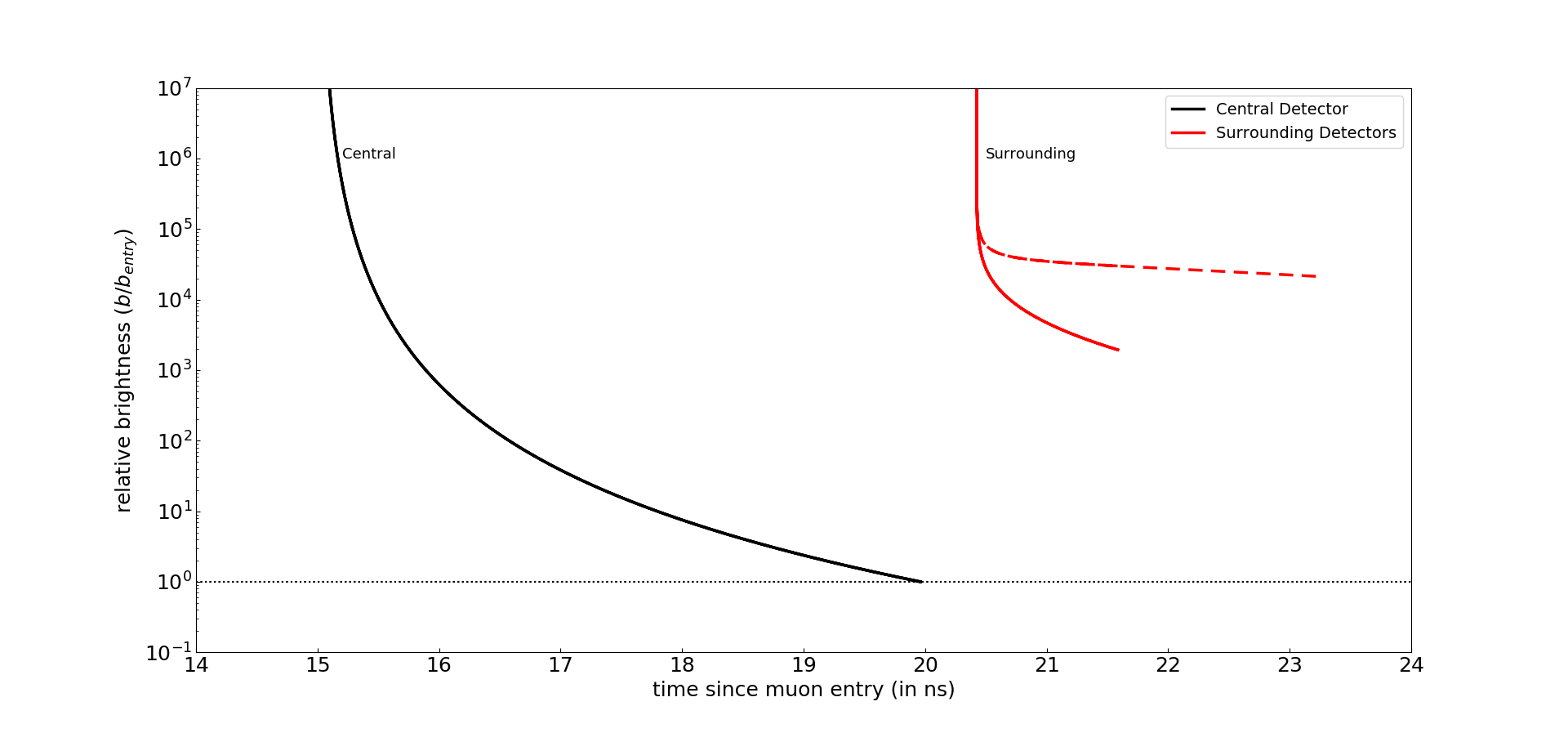}
        \caption{A plot of relative brightness versus time for the Cherenkov images observed in the case of a muon moving straight down and then striking the central detector. After observing the first RID at 20.42 ns, the three surrounding video detectors will see two images of the muon track with time, with one fainter than the other. The image going up towards the entry point will not be observable after 21.58 ns but the one going down will continue to be seen till 23.22 ns. Because the two images usually differ so much in brightness, a PMT detector will effectively measure only the light curve of the brighter of the two images.}
            \label{vertical_b_vs_t}
        \end{figure}

The third possible detector type, the time-sensitive PMT, will record only a light curve -- a plot of image brightness versus time. The light curve of the vertically descending muon (which results in a single vertically ascending Cherenkov image) is shown in Figure~\ref{vertical_b_vs_t}.

\subsection{Horizontal Incidence}

This case considers the muon travelling horizontally through the water entering the tank at point $A(x_1,\sqrt{R^2-x^2_1},h)$ and leaving through $B(x_2,\sqrt{R^2-x^2_2},h)$, where the muon incidence height $h$ can be anywhere between $0$ and $H$. In correspondence with critical height $z_C$ for a downward moving muon, a critical distance $x_C$ from entry point $A$ exists where, if $x_C < AB$, where $AB$ is the distance between $A$ and $B$, the detector will be exposed to a pair event at $x_C$. The location of $x_C$ is given by
    \begin{equation}\label{xc}
        x_C = L \cos \alpha - \frac{c_w L \sin \alpha}{\sqrt{v^2-c_w^2}}.
    \end{equation}
Of course if $x_C > AB$, the detector will perceive the Cherenkov image just moving from $A$ to $B$ as expected classically. 

When $x_C < AB$, the two images of the muon will be seen at distances $x_{\pm}$ from $A$ where $x_{\pm}$ is given by
    \begin{equation}{
        x_{\pm} =   \frac{c_w^2 t_{total} v-L v^2 \cos \alpha \pm \sqrt{v^2 \left(c_w^2 \left(L^2+t_{total}^2 v^2\right)-2 c_w^2 L t_{total} v \cos \alpha +L^2 v^2 \cos ^2\alpha-L^2 v^2\right)}}{c_w^2-v^2} }.
    \label{eq:horizontal_xpm}
    \end{equation}
Here, $x_-$ is the distance (from $A$) of that Cherenkov image which is moving towards $B$ -- along the original direction of muon track -- while $x_+$ is the distance of the Cherenkov image (from $A$) which moves towards $A$, in the opposite direction to the muon's motion. The angular locations of muon images in this case are given by $\phi_{\pm}$ where $\phi$ is the angle between the line joining the detector with the image location and the line joining the detector with the point of entry $A$. In this case, an angle $\phi=0 ^{\circ}$ denotes the detector-entry point line. 

\begin{figure}[h]
            \centering
            \hspace*{-1.0cm}
            \includegraphics[scale=0.4]{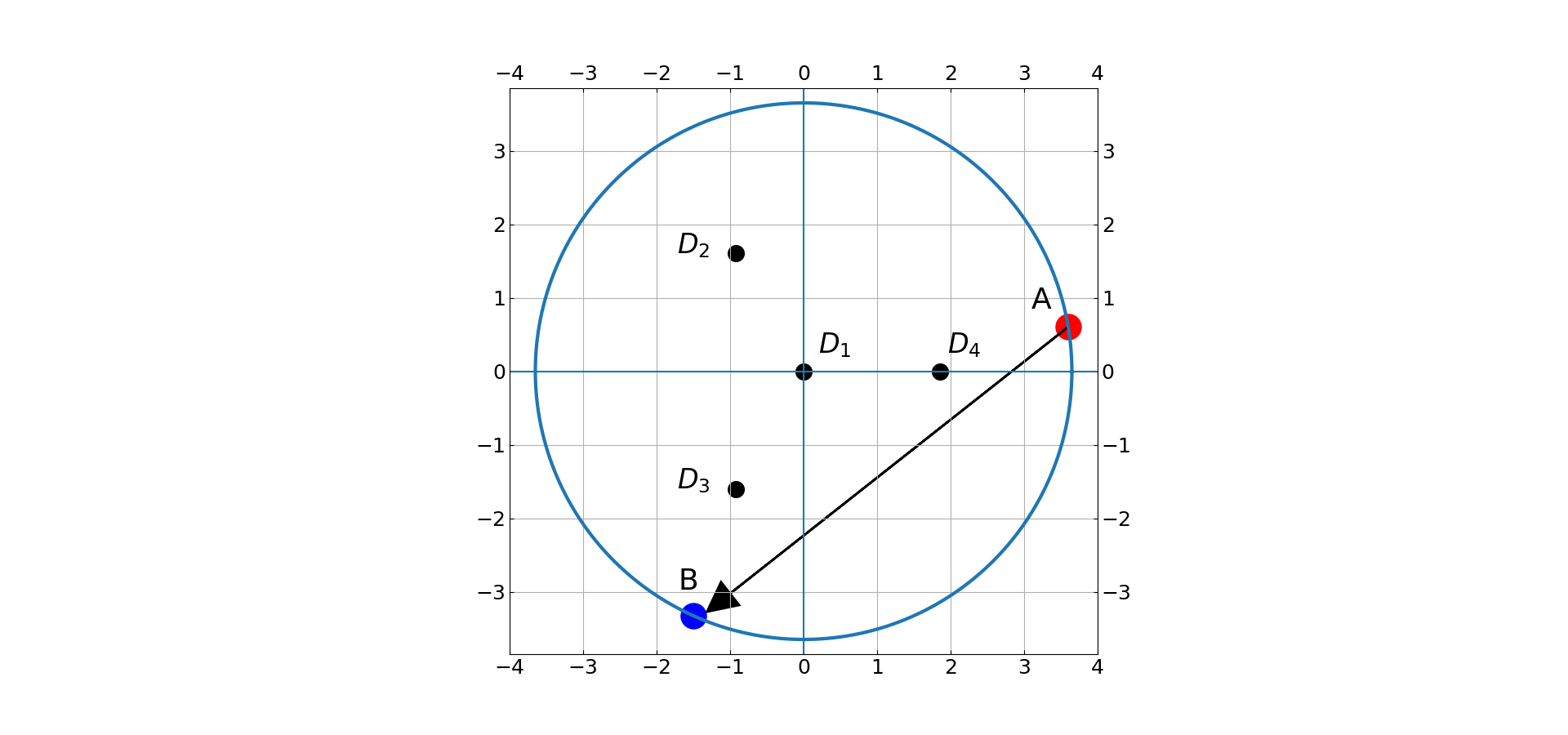}
        \caption{Top view of the WCD tank for a muon entering horizontally through point A and leaving through B. The central detector ($D_1$) and the surrounding detectors ($D_2, D_3, D_4$) are also shown.}
            \label{cross_section}
        \end{figure}

For clarity, a specific case in which the muon enters the tank at an arbitrarily chosen point $A(3.6,0.6,0.5)$ and leaves at another arbitrarily chosen point $B(-1.5,-3.3,0.5)$, is discussed here. A transverse cross-section of the tank is shown in Figure~\ref{cross_section}. 
 
\begin{figure}[h]
            \centering
            \hspace*{-1.0cm}
            \includegraphics[scale=0.4]{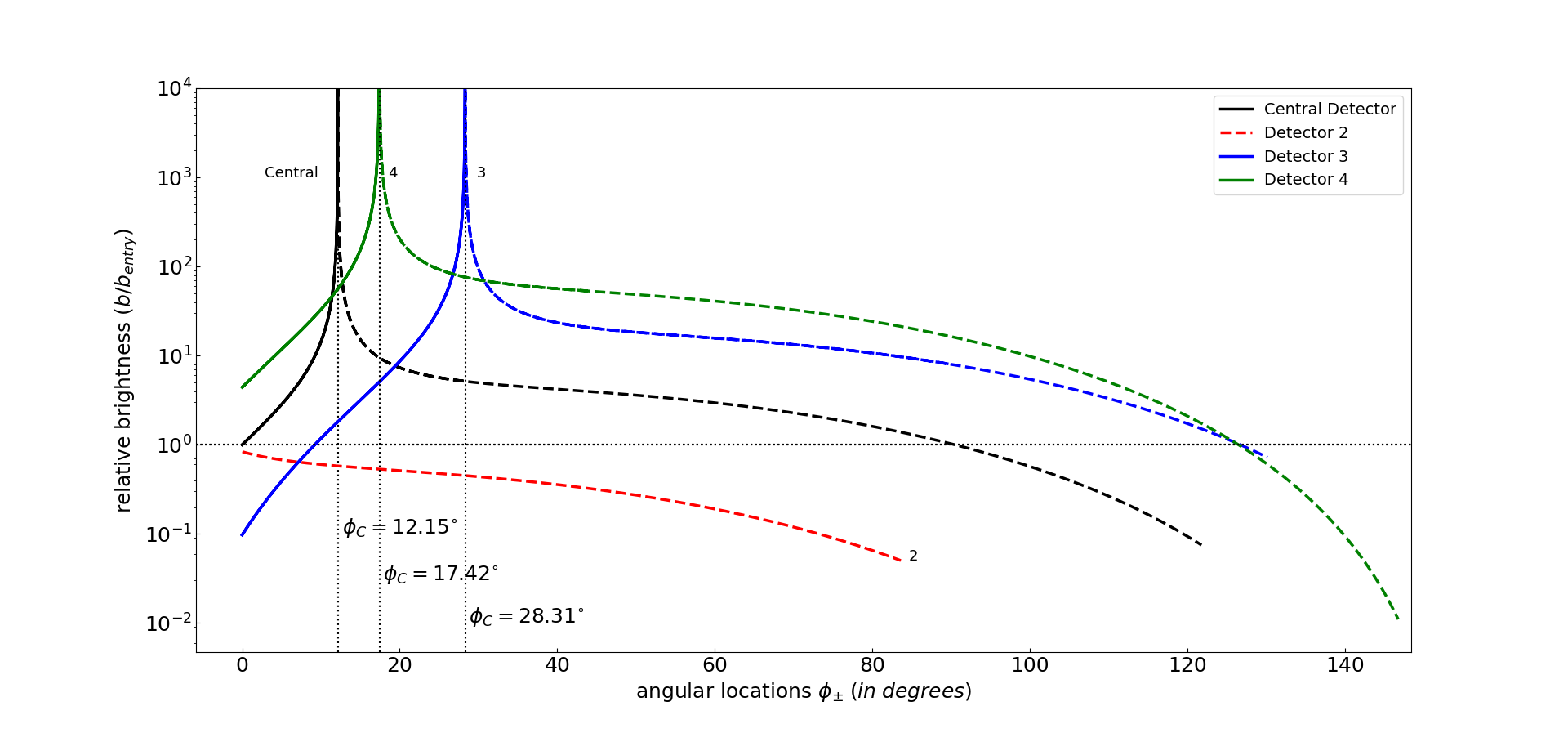}
        \caption{A plot of relative brightness vs angular locations $\phi_{\pm}$ of Cherenkov images of a muon entering the WCD horizontally. The dashed curve corresponds to the image going towards the exit point B, while the solid curve represents the image going back towards the entry point A.}
            \label{horizontal_b_vs_phi}
        \end{figure}

A graph of the brightness of Cherenkov images versus their angular locations for this case is shown in Figure~\ref{horizontal_b_vs_phi}. The brightness of the muon image at each point in its track has been normalized with respect to its brightness at the entry point $A$ as seen by the central detector. The information in this graph could be recovered from a static digital camera image; a line of variable brightness that encodes the variation of brightness of Cherenkov images with their corresponding angular locations. Note that detector 2 in this case will not see an RID, because this detector is never inside the Cherenkov cone of the muon. The other 3 detectors, however, will each record an RID event.  It can also be seen from Figure~\ref{horizontal_b_vs_phi} that none of the light curves shows a strong inverse square dependence on the distance of the Cherenkov images of the muon from the detector. This is primarily attributed to the ratio of the perpendicular distance between the detector and the muon track to the total length of the muon track. The higher this ratio is, the greater the distance dependence of the images. For a typical WCD, all of the distances tend to be of the order of a meter, whereas for a typical Imaging Atmospheric Cherenkov Telescope (IACT), the dynamic range of distances changes from meters to hundreds of kilometers, making the distance dependence of Cherenkov images relatively modest for WCDs when compared to IACTs \citep{{2020ApJ...889..122N}}. 
\break

\begin{figure}[h]
            \centering
            \hspace*{-1.0cm}
            \includegraphics[scale=0.42]{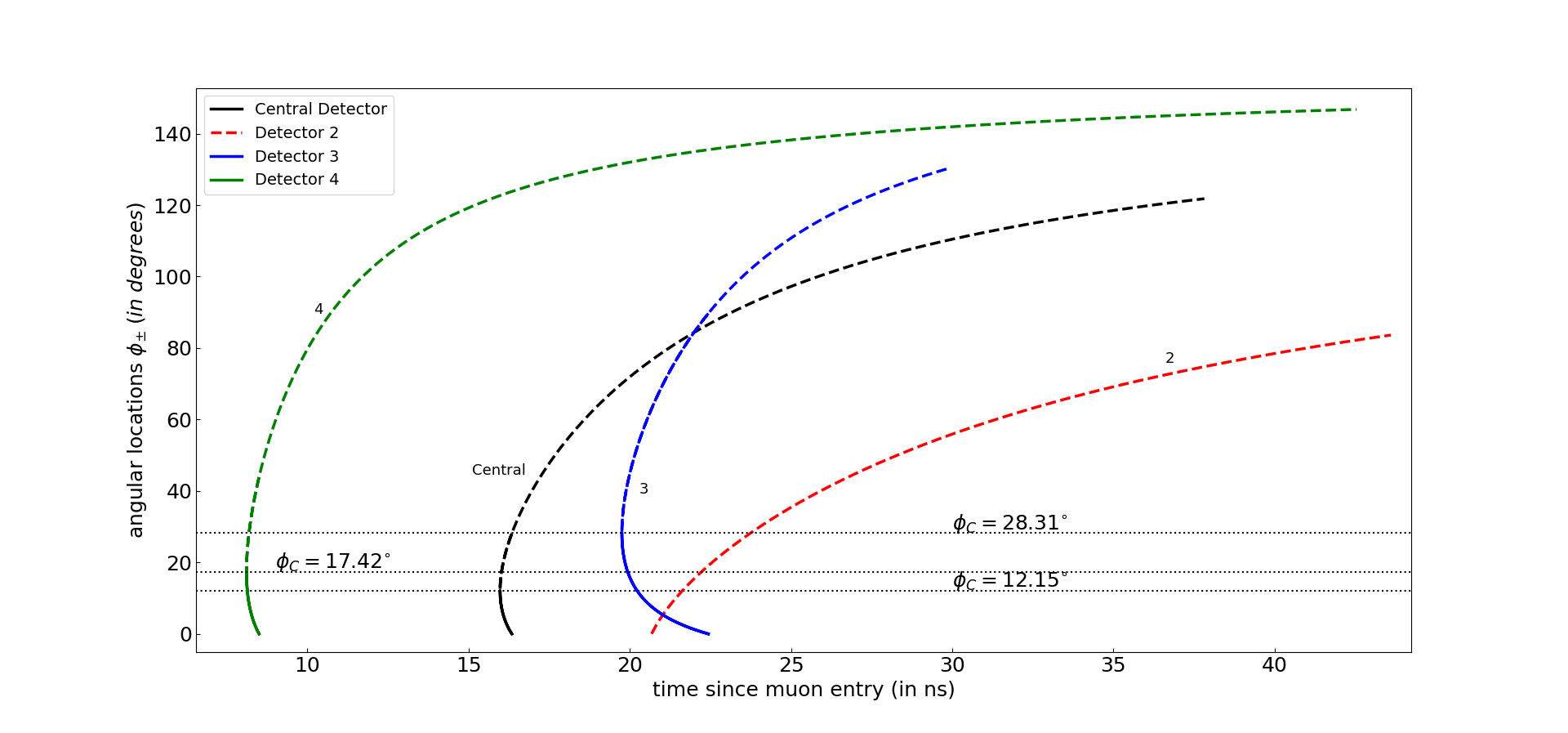}
        \caption{A plot of angular positions of images versus total time for a muon entering the water tank horizontally. The short solid curves depict Cherenkov images moving back towards the entry point A. The long dashed curves depict images moving forward towards the exit point B. The angle $\phi_C$ is the critical angle where each detector sees the first RID and it corresponds to the critical distance $x_C$. The figure also shows the relative closeness of the first RID locations ($x_C$) to the entry point. The shorter the solid curve, the closer $x_C$ is to the point of entry of muon.}
            \label{horizontal_phi_vs_t}
        \end{figure}

\begin{figure}[h]
            \centering
            \hspace*{-1.0cm}
            \includegraphics[scale=0.42]{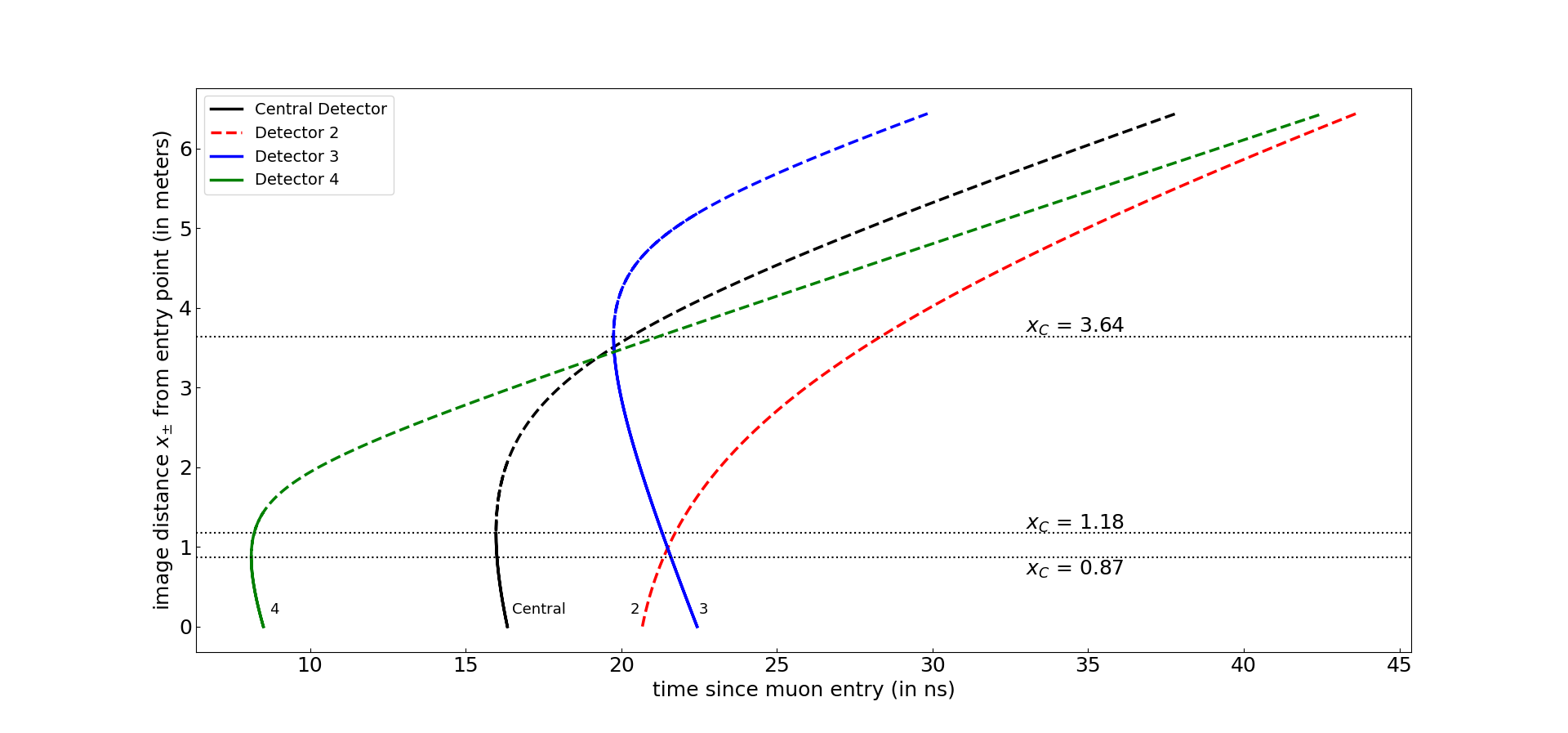}
        \caption{A plot of the distances of the Cherenkov images of the muon from the entry point $A$ versus the total time since the muon entered the tank horizontally.}
            \label{horizontal_x_vs_t}
        \end{figure}

\begin{figure}[h]
            \centering
            \hspace*{-1.0cm}
            \includegraphics[scale=0.4]{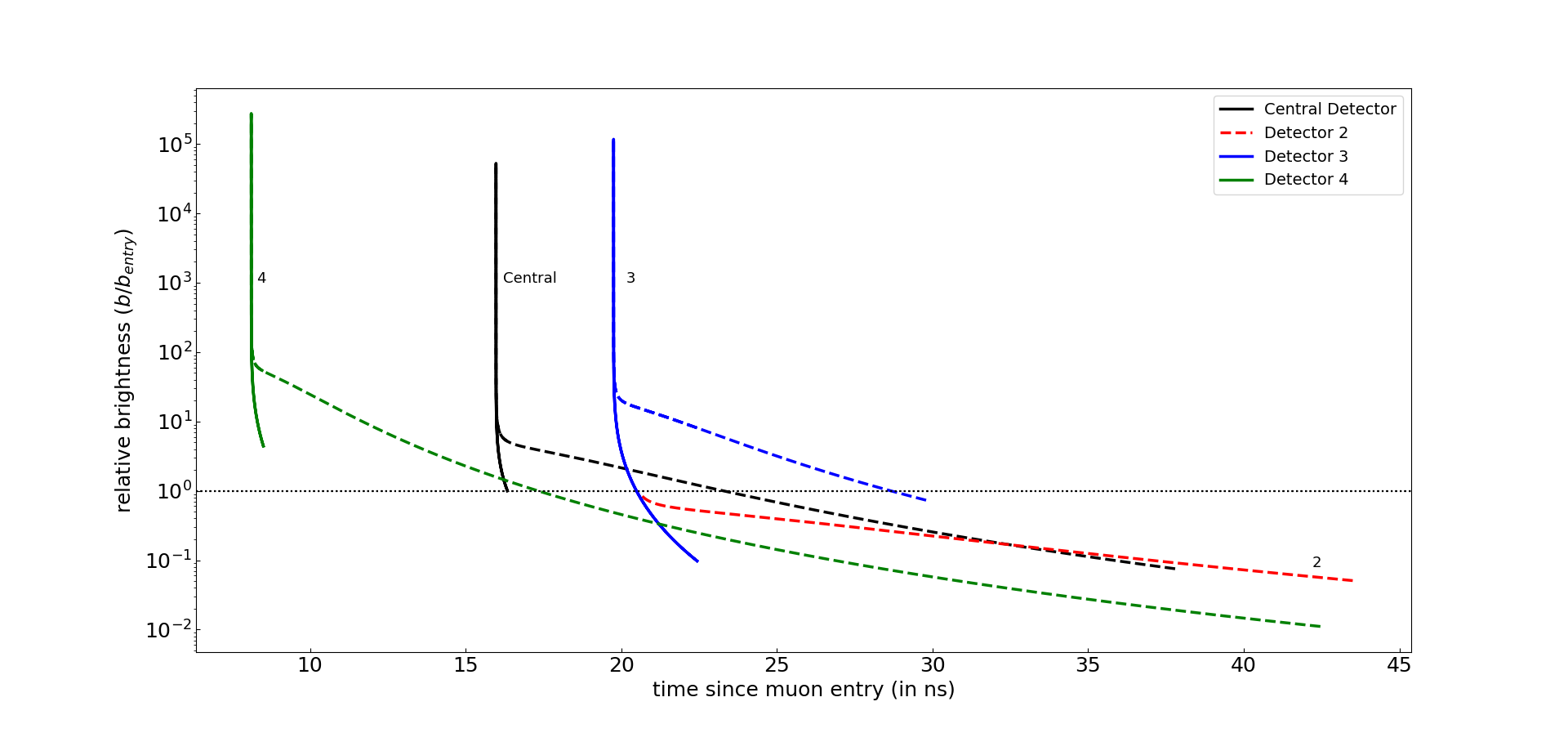}
        \caption{A plot of relative image brightness versus time for a muon entering the WCD horizontally. Note that the Cherenkov images of the same muon track will appear to different detectors at different times with different relative brightness during its entire path inside the WCD. At 22 ns, in total five Cherenkov images of different brightness will be visible collectively to all the detectors.}
            \label{horizontal_b_vs_t}
        \end{figure}

A video detector in this case will see the development of Cherenkov images along the muon track with time. Data extracted from the resulting video would show the angular positions of the images with time as shown in Figure~\ref{horizontal_phi_vs_t}. A graph of the distances of the Cherenkov images of muon from its point of entry versus total time is shown in Figure~\ref{horizontal_x_vs_t}. The information on this graph would only be discernible from a high-speed video detector. A PMT in this case will record only the light curve shown in Figure~\ref{horizontal_b_vs_t}.

From inspection of Figure~\ref{horizontal_b_vs_t}, it is clear that the first RID pair of Cherenkov images appears with high brightness (formally infinite) and fades quickly. The two images disappear at different times with respect to the different detectors. The one detector in this example that does not see an RID never sees the single Cherenkov image appear as bright as its pair-counterparts, as observed by other detectors. Also, the solo Cherenkov image this detector sees moves monotonically from $A$ to $B$.

\subsection{Incidence at an Arbitrary Angle}

More realistically, consider a muon entering the WCD at some point $A$ on the top and leaving the bottom through a different point $B$. As delineated in Eq.~\ref{zc}, the muon's track makes an angle $\theta$ with respect to the vertical. 

The muon travels in a straight line through the WCD at speed $v$. The ground point $B$ may or may not be inside the tank, but the muon trajectory itself will always have the same $\theta$ as it had at point $A$. The muon will trigger Cherenkov radiation when in the tank. Each detector inside the WCD will see this Cherenkov radiation but is not guaranteed to see an RID. The recording of RID by any detector depends on a variety of parameters including but not limited to $\theta$, $L$ and $\alpha$ as indicated in Eq.~\ref{zc}. The apparent heights of the Cherenkov images are given by Eq.~\ref{zpm}. 

\begin{figure}[t]
            \centering
            \hspace*{-1.0cm}
            \includegraphics[scale=0.4]{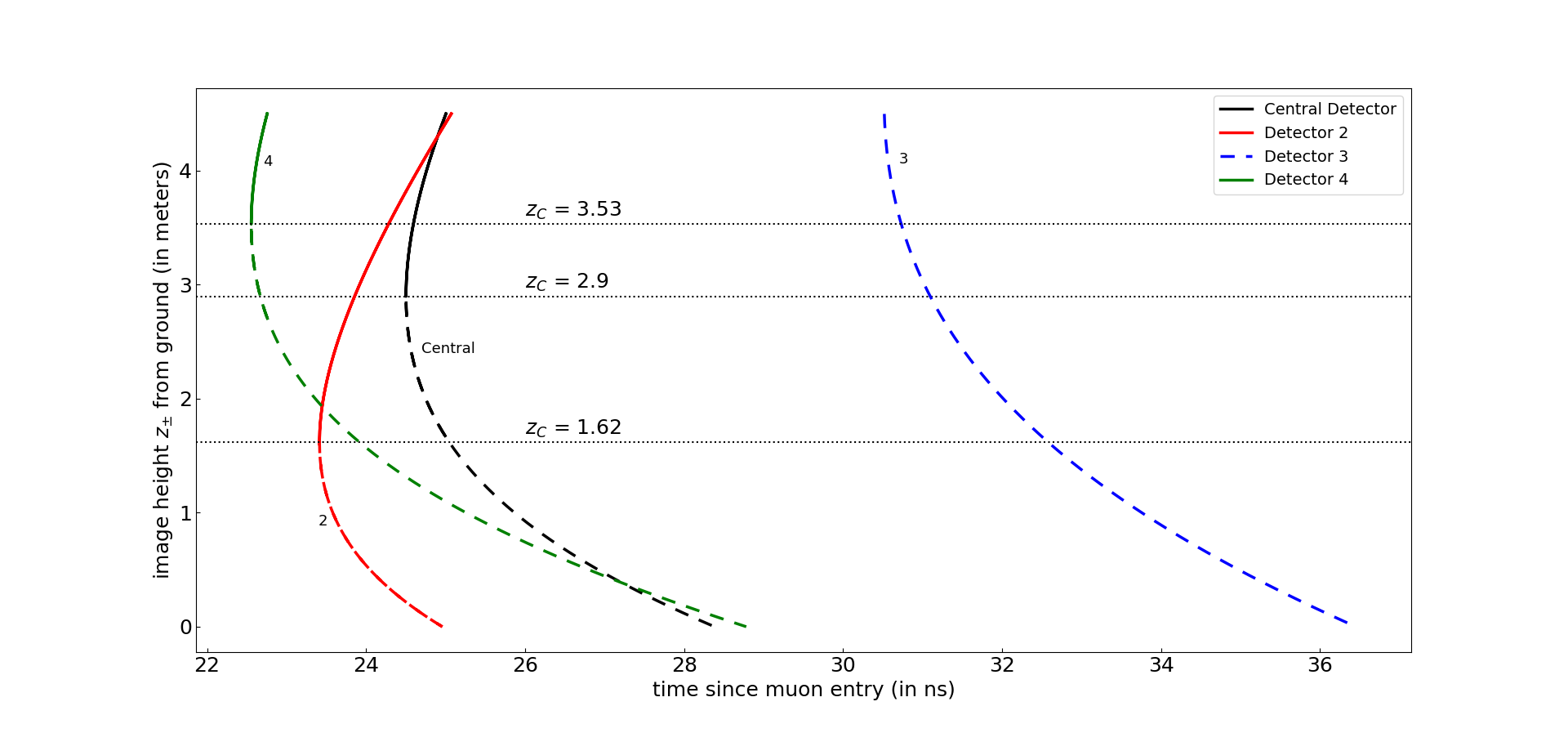}
        \caption{A plot of image heights versus time for a muon entering the tank with the example oblique trajectory given in the text. A solid curve represents the Cherenkov image moving upwards while a dashed curve means that the image is moving downwards towards the bottom. The central detector, as well as detectors 2 and 4, all see an RID, but observe it at different heights from the ground and at different times. Detector 3, however, does not see an RID.}
            \label{oblique_z_vs_t}
        \end{figure}

We now consider a very specific example chosen to highlight generic features of RID in WCDs. In this example, the muon (or any charged particle) enters the top of the tank at $A = (2.4,2.4,H)$ with $\theta=20 ^{\circ}$ and strikes the ground at $B = (0.79, 2.68, 0)$ which is inside the tank. A graph of image heights from ground $z_{\pm}$ versus the total time $t_{total}$ is shown in Figure~\ref{oblique_z_vs_t}.

Inspection of Figure~\ref{oblique_z_vs_t} shows a time offset of approximately 22 ns between when the muon first enters the WCD at point $A$ and when any of the detectors records Cherenkov light from the muon. This time offset is due to the muon travel time from $A$ to the location of the first detected RID plus the Cherenkov light travel time from this RID location to the detector. As all detectors are at different distances from both $A$ and $B$, the time at which each detector sees an RID, is different.

\begin{figure}[h]
            \centering
            \hspace*{-1.0cm}
            \includegraphics[scale=0.35]{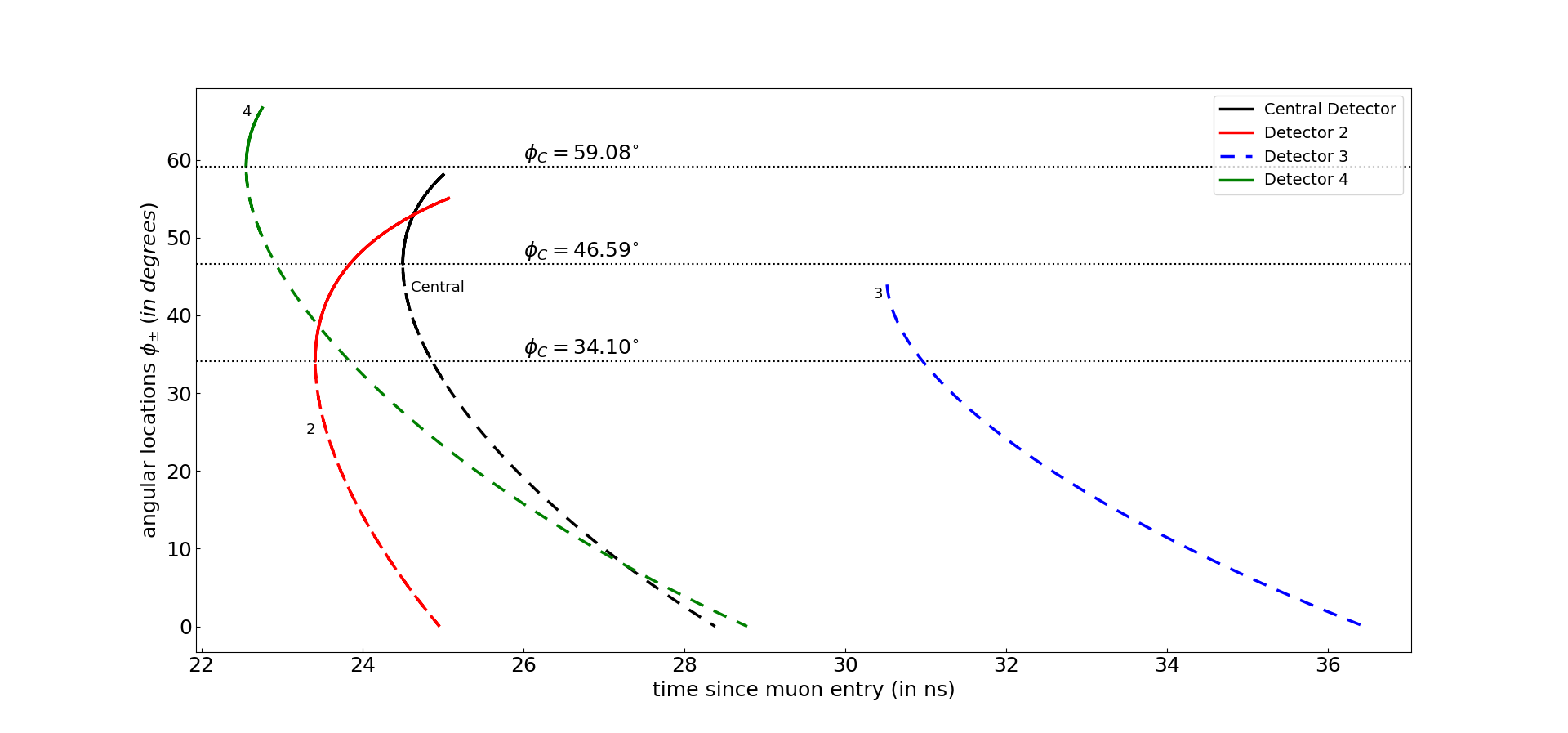}
        \caption{A plot of angular location of Cherenkov images versus total time for a muon entering the WCD at the example oblique trajectory given in the text. Note again that detector 3 does not observe any RID so that the only curve for detector 3 is the dashed one i.e. this detector will see the muon image first at 30.52 ns and last at 36.45 ns.}
            \label{oblique_phi_vs_t}
        \end{figure}

A graph of angular locations $\phi_{\pm}$ of images versus time $t_{total}$ is shown in Figure~\ref{oblique_phi_vs_t}.
Inspection of Figure~\ref{oblique_phi_vs_t} shows that $t_{total}$ is frequently double-valued with respect to $\phi$. For detectors having double-valued curves, the muon is seen for the first time when $z = z_C$ i.e. when the first RID is seen. An RID thus produces the first light observed by the detector. Also note that only detector 3 sees a Cherenkov image move classically from point $A$ to point $B$, but  only after the other three detectors have stopped observing it.

\begin{figure}[h]
            \centering
            \hspace*{-1.0cm}
            \includegraphics[scale=0.4]{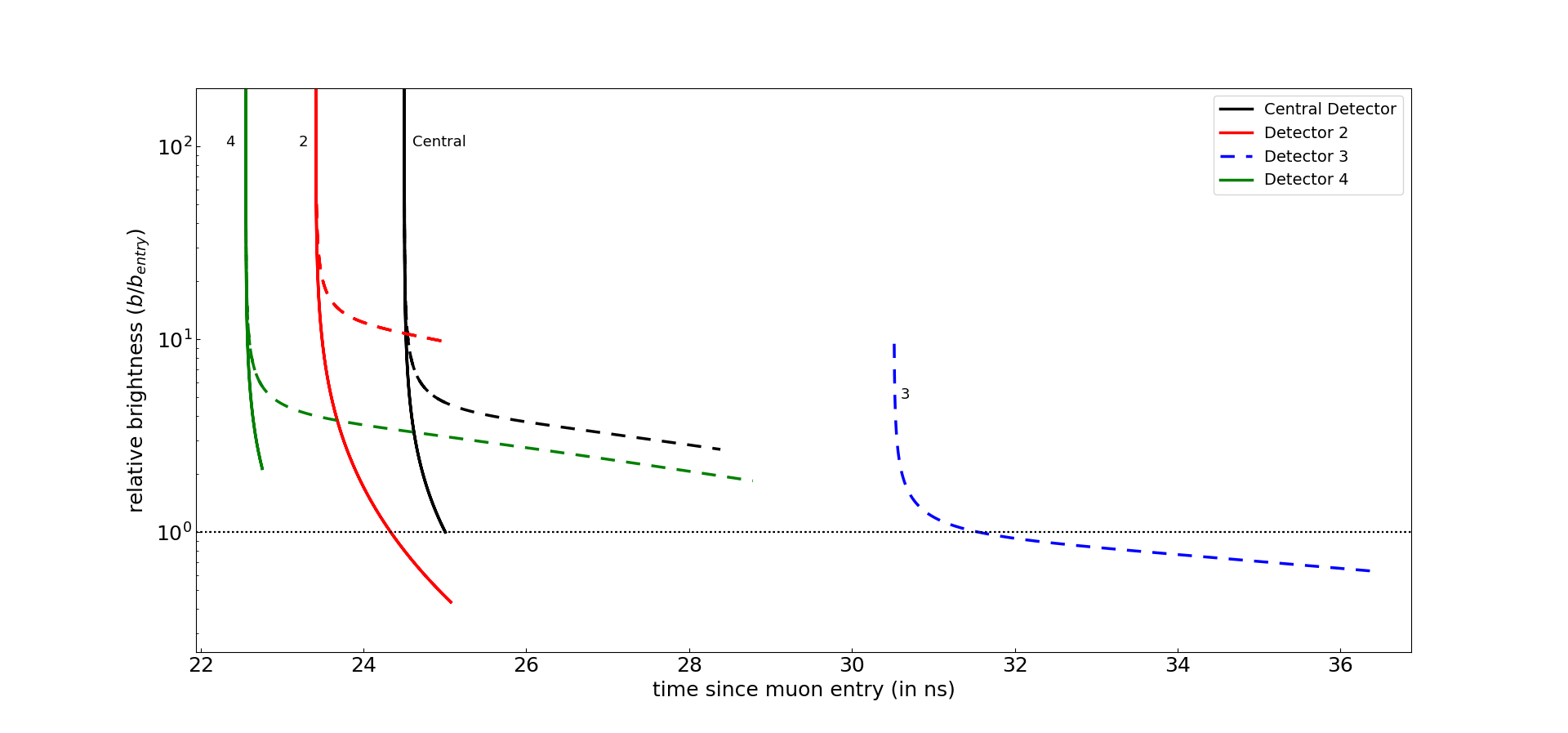}
        \caption{A plot of relative Cherenkov image brightness versus total time for a muon entering the WCD at the example oblique angle. Note that detector 3 starts measuring the brightness of the single Cherenkov image it sees only after the Cherenkov images recorded by the other three detectors have disappeared.}
            \label{oblique_b_vs_t}
        \end{figure}

A graph of relative image brightness versus time for all the four detectors is shown in Figure~\ref{oblique_b_vs_t}. From this plot, it can be deconvolved what a PMT will record. Figure~\ref{oblique_b_vs_t} shows that at roughly 25 ns after the muon enters the tank, three out of four video detectors will collectively observe five images of the same muon as opposed to the conventional notion of observing three images, one for each detector. Each of these five images will stop appearing at different times.

\begin{figure}[t]
            \centering
            \hspace*{-1.0cm}
            \includegraphics[scale=0.4]{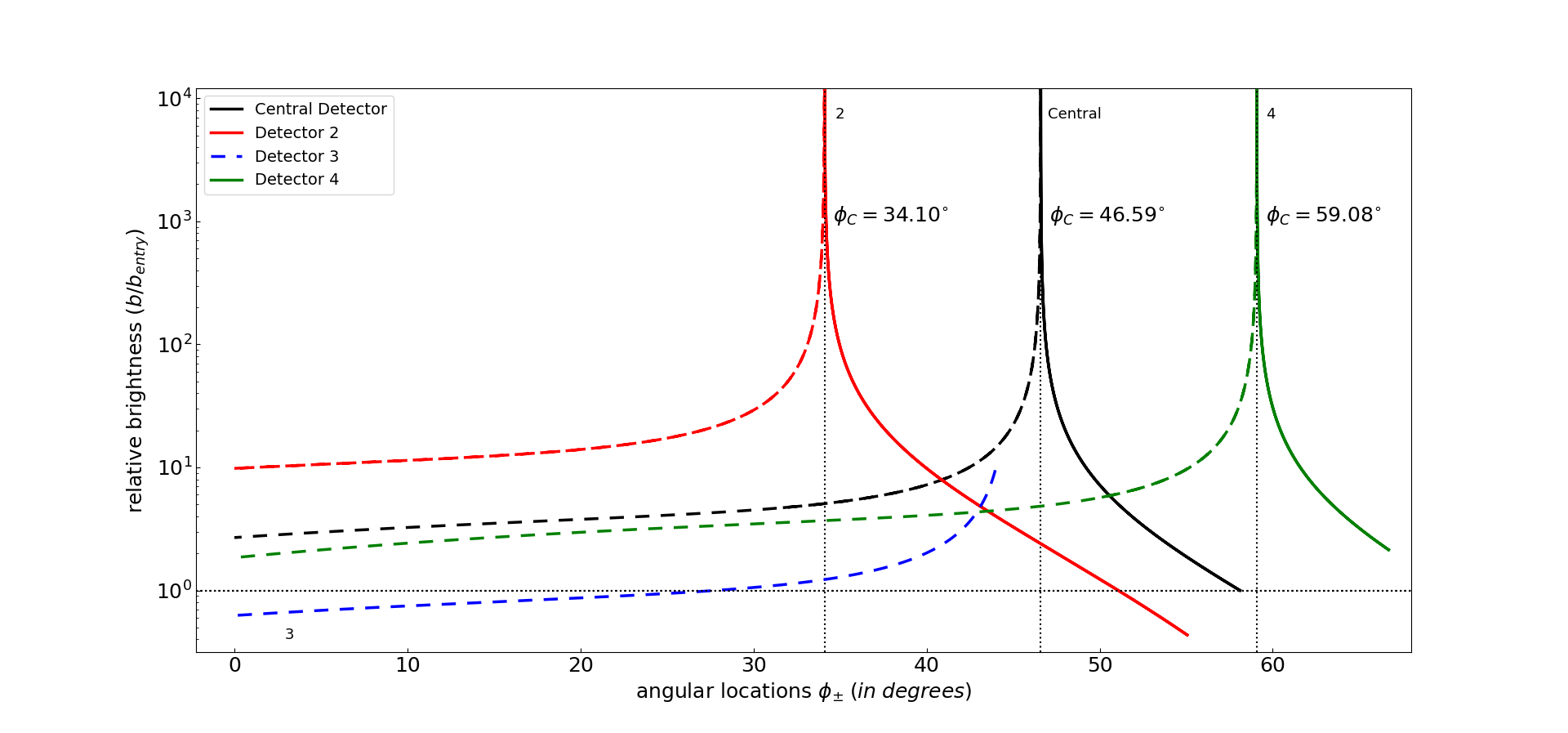}
        \caption{A plot of relative brightness versus angular locations of Cherenkov images of the muon entering the tank at the example oblique angle. An angle $\theta=0^{\circ}$ here corresponds to the detector looking directly towards the entry point $A$.}
            \label{oblique_b_vs_phi}
        \end{figure}

Finally, a graph of relative brightness vs angular locations as observed by all the four detectors is shown in Figure~\ref{oblique_b_vs_phi}. For the detectors that see an RID, the graph shows sharp peaks in the brightness of each image when $z = z_c$, which corresponds to $v_r = c_{w}$. These peaks correspond to when the Cherenkov images appear for the very first time to each detector. The top Cherenkov image disappears when it reaches the upper lid of the tank at $A$, while the bottom image disappears when it hits the ground at $B$.

\section{Monte Carlo Simulation: Fraction of Muons Showing an RID Event in a HAWC-Like Detector}

Monte Carlo simulation were performed to find the relative fraction of charged, relativistic particles entering a HAWC-like WCD that can cause an RID effect visible to at least one of the internal detectors. Two different algorithms were employed to find this fraction, with both returning similar results. In both algorithms, a cylindrical WCD tank was assumed having radius $R$ and height $H$. 

The first algorithm was written in FORTRAN and is shown in Appendix A.1. In this algorithm, muons were assumed to pass through the ground at a random location within three $R$ of the center of the cylindrical WCD. This location, labeled $B$, was then associated with a randomly chosen azimuthal angle $\phi$ relative to an $x$-axis connecting to the center of the bottom of the WCD, and then associated with a randomly chosen zenith angle $\theta$. Given $R$, $\phi$ and $\theta$ for ground location $B$, the muon's three-dimensional path was uniquely determined. Paths that did not intersect the WCD were discarded. Paths that did intersect the WCD were then assumed to model the muon moving with a downward component through the WCD. The location $A$ where the muon entered the WCD was found. The Cherenkov cone angle $\beta$ was extended from $A$ downward to see if any detectors were inside. The fraction of interior paths to total paths was then computed.

A second algorithm was independently conceived and coded in C++ by the other author and is shown in Appendix A.2. In this algorithm, muons were assumed to come from all possible directions. A random entry point $A$ was first chosen to be anywhere on the top or walls of the tank. Next, a random exit point $B$ was chosen to be located anywhere on the walls or the bottom of the tank. Then a muon track was constructed passing through these points and going down (The $z$-coordinate of exit point was always less than or equal to the $z$-coordinate of entry point). The distance from the entry point $A$ where the first RID would be seen by at least one of the four HAWC-like PMT detectors was found using equation~\ref{xc} for that muon track. If that RID location was inside the tank, that particular detector was considered to be showing an RID and that muon track was not checked further for any RIDs by the remaining detectors. This process was repeated for a large number of muon tracks which were then subsequently checked for an RID event by at least one of the four detectors.

The two algorithms were run for $1$ billion muon tracks using parameters describing a HAWC-like WCD as shown in Fig.~\ref{figure_hawc_system} and Fig.~\ref{cross_section}. Both algorithms found that of all the muons entering the WCD, between $85$ to $90\%$ would create an RID as seen by at least one detector. This indicates that RID events should be seen in HAWC-like WCDs quite commonly. These codes are available on GitHub \footnote{\url{https://github.com/neeravkaushal/RIDs-in-WCDs/tree/v1.0}} under an MIT License and version 1.0 is archived in Astrophysics Source Code Library (ASCL) \citep{ASCL}.

\break
\section{Discussion} \label{sec:Discussion}

Relativistic Image Doubling (RID) is a simple concept in optics where, as opposed to physical or gravitational lenses, relativistic speed itself creates multiple images of the source \citep{2018AnP...53000333N}. RID is an observer effect -- the muon does not split, for example -- and RID can only be observed when something moves faster than light in a given medium. Surprisingly, perhaps, RID effects are actually quite common on Earth, but typically occur on such a short time scale that they are not noticed. To date, only one RID effect has been reported \citep{2016SciA....2E1691C}, although recently many more have been hypothesized (see, for example, \citet{2020ApJ...889..122N}).

A common medium where RIDs may be observed is water. Many ground-based high-energy gamma-ray observatories use water to facilitate the creation of Cherenkov radiation caused (liberated) by superluminal particles. However, to the best of our knowledge, none of the science teams that run these experiments have published an analysis that acknowledges the RID effect.

RID has several uniquely defining characteristics. One is that, for a given detector, before an RID is detected, no Cherenkov light at all will be seen. It is not that the Cherenkov images are dim, it is that they are not there. For any one detector, the RID appears suddenly and exactly in the direction of the Cherenkov cone of the descending muon. The line connecting the Cherenkov cone to the detector gives the direction of the RID as seen by the detector. Another defining characteristic is that the RID event, as seen by any given detector, is very bright. This high brightness is recorded even though the Cherenkov track is uniformly bright along its track. The reason the RID appears bright is because the detector receives light from a long section of the uniformly bright track all in a short period of time. 

After a video detector records an RID event, two Cherenkov images from the RID could be seen moving along the track of the muon, but in opposite directions. The two images will start with formally infinite angular speed, which drops quickly after appearance. These images will start with the same great brightness, but both will quickly fade, and their relative brightness will diverge from unity. This attribute can be important for identifying and verifying images resulting from RID. In the example case of a horizontally moving muon, this fading is shown in Figure~\ref{horizontal_b_vs_t}.

One of the two Cherenkov images will eventually disappear at the location where the actual muon exited the WCD, while the other Cherenkov image will suddenly disappear at the location where the muon {\it entered}. It is interesting to note that the backwards-moving Cherenkov image of the muon is actually observed time-backwards by the detector. Although each Cherenkov image will usually appear to fade as it goes, sometimes a competing brightening effect occurs when an image nears a detector.

One might argue that RID is a trivial effect. Who cares that two images of the same muon track are sometimes visible simultaneously? In our view, as also noted in \citep{2020ApJ...889..122N}, RID is a novel optical effect caused not by lenses but solely by relativistic kinematics -- and so is interesting basic physics even without a demonstrated usefulness. However, discovering and tracking RID effects in WCDs may prove useful. Such observations could add information that better allow muon track locations and orientations to be recovered, or give an independent method of confirming the muon's trajectory. Studying the basic science of other multiple-imaging optical effects is common practice, for example in gravitational lensing \citep{1987PhDT........12N, 1992grle.book.....S} and temperature inversions effects in the Earth's atmosphere \citep{1999YoungsFamousWebPages}. 

Why have RID effects remained so obscure? One reason is that they have no classical analog -- they depend crucially on the finiteness of the speed of light. Classical thinking may allow scientists to accurately visualize how the muon actually moves through the WCD, but not always how it appears to the detector \citep{2020ApJ...889..122N}. 

RID effects may be thought of as immeasurable by some, because both images fade within nanoseconds, as shown in Figure~\ref{oblique_b_vs_t}. However, tracking the images for even a short angular distance would be interesting. As computer technology and miniaturization is increasing the frame rate that can be captured, imagers appearing over the past few years are becoming capable of recording sub-nanosecond events \citep{2016SciA....2E1691C}. These technological innovations raise the possibility of placing video detectors inside WCDs that can resolve RID events in both time and angle. Alternatively, simple digital cameras may be placed that can resolve the Cherenkov images only in angle, leaving the temporal resolution to the PMTs.

There are also attributes of realistic detectors that may obscure RID effects. One is that the detector itself is not just a point -- it may be so large that light travel time across its surface is significant when compared to $t_{light}$. Then any light curve that a WCD PMT measures will convolve the size and shape of the PMT, not just geometry inherent to the muon's path. 

Because RID effects will look different to different detectors even for the same muon event, the locations of individual detectors are paramount. Therefore, simply adding together the brightness of different images from multiple PMTs, at the times of the brightness measurements, for example, will typically convolute RID effects beyond recognition. However, a careful reconstruction accounting for the timing of separate RID events as seen by different detectors should be possible that could enhance RID detection and better determine the muon's real track inside the WCD.

There are many interesting RID and detector-related effects that were deemed too complex to be incorporated in this primarily conceptual work. One such effect is the deceleration of the muon as it moves downward through the WCD. A muon that slows significantly will force $z_C$ to occur higher in the tank. Additionally, the size of the detectors may cause a convolution of measurements over the time it takes for light to cross the detector. A full treatment of these effects would likely require more detailed and complex computer modeling, but not change the character of the reported results. 

In sum, resolving relativistic image doubling events in time and space would recover a novel type of optics caused not by lenses but by relativistic kinematics alone. Further, tracking simultaneous images as they emerge and diverge in angle, angular speed, and relative brightness could resolve or independently confirm information about the muon's trajectory, including the brightness along its path as described through equations in Section 2.

\section*{Acknowledgments} \label{sec:A}
We thank Cameron Shock, Rishi Babu and an anonymous referee for insights.


\appendix

As a check on the accuracy of the results, the simulations were computed in different programming languages. NK's Python and C++ codes are listed first followed by RJN's FORTRAN code. 

\definecolor{codegreen}{rgb}{0,0.6,0}
\definecolor{codegray}{rgb}{0.5,0.5,0.5}
\definecolor{codepurple}{rgb}{0.58,0,0.82}
\definecolor{backcolour}{rgb}{0.95,0.95,0.92}

\lstdefinestyle{mystyle}{
    backgroundcolor=\color{backcolour},   
    commentstyle=\color{codegreen},
    keywordstyle=\color{magenta},
    numberstyle=\tiny\color{codegray},
    stringstyle=\color{codepurple},
    basicstyle=\ttfamily\footnotesize,
    breakatwhitespace=false,         
    breaklines=true,                 
    captionpos=b,                    
    keepspaces=true,                 
    numbers=left,                    
    numbersep=5pt,                  
    showspaces=false,                
    showstringspaces=false,
    showtabs=false,                  
    tabsize=2
}

\subsection{FORTRAN Code: Monte Carlo Simulation for fraction of muons showing an RID event}
\lstset{style=mystyle}
\begin{lstlisting}[language=Fortran]

c Computes fraction of HAWC events that show RID.
       program fractionRID
c  
c      integer*4, parameter :: n=10
       integer*4 i, ntot, nRID, n, nmax
       real*8 c, pi, seed, pin, eps, Cangle
       real*8 R, H, Rd
       real*8 xd(4), yd(4), zd(4), th(4)
       real*8 xg, yg, zg, rg
       real*8 xs, ys, zs, rs
       real*8 xm, ym, zm, rm
       real*8 xe, ye, ze, re
       real*8 x_ge, y_ge, z_ge
       real*8 x_de(4), y_de(4), z_de(4), r_de(4)
       real*8 top, bot
       character*3 before, now, RID
c                             Constants
       nmax=1000000000
       c=3.0d8
       pi=dacos(-1.0d0)
       seed=2.8d0
       eps=0.1d0
       Cangle=dacos(1.0d0/1.33d0)*(180.0d0/pi)
       write (*,*) ' Input seed, please: '
       read (*,*) seed
c                             R=radius of tank (meters)
c                             H=height of tank (m)
c Source: https://www.sciencedirect.com/
c science/article/pii/S0273117713001695?via%3Dihub
       R=7.30d0/2.0d0
       H=4.5d0
c                             Ground dist. to detectors (check)
       Rd=7.3d0/4.0d0
       Rd=1.850d0
c                             Location of four ground detectors
       xd(1)=0.0d0
       yd(1)=0.0d0
       zd(1)=0.0d0
       xd(2)=-Rd
       yd(2)=0.0d0
       zd(2)=0.0d0
       xd(3)=Rd*cos(pi/3.0d0)
       yd(3)=Rd*sin(pi/3.0d0)
       zd(3)=0.0d0
       xd(4)=Rd*cos(pi/3.0d0)
       yd(4)=-Rd*sin(pi/3.00)
       zd(4)=0.0d0
c
       ntot=0
       nRID=0
c ********************
c                             Grand loop over muons
c ********************
       do 700 n=1, nmax
c                             Location muon hits ground
       xg=3.0d0*R*(2.0d0*pin(seed) - 1.0d0)
       yg=3.0d0*R*(2.0d0*pin(seed) - 1.0d0)
       zg=0.0d0
       rg=sqrt(xg**2 + yg**2)
       if (rg.ge.5.0d0*R) goto 700
c                             Direction muon arrived on sky
 200   xs=(2.0d0*pin(seed) - 1.0d0)
       ys=(2.0d0*pin(seed) - 1.0d0)
       zs=pin(seed)
       phi=atan(ys,xs)
c                             Angular direction of muon
       phi=2.0d0*pi*pin(seed)
       theta=acos(zs)
c                             Horizon dimming
       if (pin(seed).gt.zs**2) goto 200
c
c ********************** Find where muon enters tank ***
c
       xm=xg
       ym=yg
       zm=zg
       rm=sqrt(xm**2 + ym**2)
c                             Is muon in the tank?
       now='out'
       if (rm.lt.R.and.zm.lt.H) now='in'
c                             Step the photon back along track
 300   xm=xm + xs*eps
       ym=ym + ys*eps
       zm=zm + zs*eps
       rm=sqrt(xm**2 + ym**2)
c      write (*,310) xm/R, ym/R, zm/H, rm/R
c310   format (2x, 4(f7.4, 3x))
c                             If muon missed tank try again
c      if (rm.gt.3.0d0*R) write (*,*) ' Muon missed tank: wide'
       if (rm.gt.3.0d0*R) goto 700
c      if (zm.gt.H) write (*,*) ' Muon missed tank: high'
       if (zm.gt.H) goto 700
       before=now
c
       now='out'
       if (rm.lt.R.and.zm.lt.H) now='in'
c      if (before.eq.'out'.and.now.eq.'in') write (*,*) "In"
c                             If muon left tank then analyze
c      if (before.eq.'in'.and.now.eq.'out') write (*,*) "Out"
       if (before.eq.'in'.and.now.eq.'out') goto 500
       goto 300

c                             The muon has just entered the tank
c                             Three points: g, e, d
c                             For ground, entry, detector
c                             Find the vertex angles
 500   xe=xm
       ye=ym
       ze=zm
       re=rm
c      write (*,*) 
c      write (*,*) ' xm / R = ', xm/R
c      write (*,*) ' ym / R = ', ym/R
c      write (*,*) ' zm / H = ', zm/H
c      write (*,*) ' rm / R = ', rm/R
c      write (*,*) ' rg / R = ', rg/R
c
       x_ge=xg - xe
       y_ge=yg - ye
       z_ge=zg - ze
       r_ge=sqrt(x_ge**2 + y_ge**2 + z_ge**2)
c      write (*,*) '   phi = ', phi*(180.0d0/pi)
c      write (*,*) ' theta = ', theta*(180.0d0/pi)
c      write (*,*) 
c
       RID='no'
       ntot=ntot + 1
       do 600 i=1, 4
       x_de(i)=xd(i) - xe
       y_de(i)=yd(i) - ye
       z_de(i)=zd(i) - ze
       r_de(i)=sqrt(x_de(i)**2 + y_de(i)**2 + z_de(i)**2)
c 
       top=x_ge*x_de(i) + y_ge*y_de(i) + z_ge*z_de(i)
       bot=r_ge * r_de(i)
       th(i)=acos(top/bot)*(180.0d0/pi)
c      write (*,*) ' th(i) = ', th(i)
       if (th(i).lt.Cangle) RID='yes'
 600   continue
       if (RID.eq.'yes') nRID=nRID + 1
 700   continue
c
       write (*,*) ' nRID, ntot = ', nRID, ntot
       write (*,*) ' RID fraction = ', float(nRID)/float(ntot)
       
c                             Stop the madness.
 999   stop
       end
c
c*******************************************************
c
c  Subroutine that uses pi**n algorithm.
       real*8 function pin(pseed)
       implicit real*8 (a-h,o-z)
c
       pi=dacos(-1.0d0)
       pseed=pseed*pi
       if (pseed.gt.10.0d0) pseed=pseed/10.0d0
       pin=pseed*10000.0d0 - int(pseed*10000.0d0)
c
       return
       end


\end{lstlisting}
\break

\subsection{PYTHON Code: Simulation for horizontal incidence of muon}

\lstset{style=mystyle}
\begin{lstlisting}[language=Python]


#!/usr/bin/env python
#coding: utf-8

#Import Libraries
import matplotlib.pyplot as plt
from numpy import sin, cos, sqrt, pi, linspace, arange, deg2rad, rad2deg, array, zeros
from numpy import arcsin, arccos,sort, argsort, argwhere, argmin, argmax, interp, concatenate
from scipy.spatial import distance
import warnings
from numpy import linalg as LA
warnings.simplefilter('ignore')

#Initialize Parameters 
n      = 1.33 #------------------------------------ Refractive index of medium
c      = 299792458/n #----------------------------- Speed of light in medium
R      = 7.3/2 #----------------------------------- Radius of tank
v      = n * c #----------------------------------- Particle Speed
times  = linspace(1e-11,1e-7,200000)
c1     = (0, 0, 0)  #------------------------------ Central    PMT number 1 
c2     = (1.85*cos(2*pi/3) , 1.85*sin(2*pi/3), 0) # Non-Radial PMT number 2
c3     = (1.85*cos(4*pi/3) , 1.85*sin(4*pi/3), 0) # Non-Radial PMT number 3
c4     = (1.85*cos(0     ) , 1.85*sin(0     ), 0) # Radial     PMT number 4
xA,xB,h= 3.6,-1.5, 0.5
A      = array( [xA ,   sqrt(R**2-xA**2), h] ) #--- Entry Point of muon
B      = array( [xB ,  -sqrt(R**2-xB**2), h] ) #--- Exit point of muon
AB     = B-A
nAB    = LA.norm(AB) #-----------------------------Muon path length
den    = c*c - v*v
print('A : ', A)
print('B : ', B)


#Bird's View of tank  
plt.figure(figsize=(10,8))
angs   = linspace(0,6.28,1000)
xs,ys  = R*cos(angs), R*sin(angs)
plt.plot(xs,ys,lw=3)
plt.scatter(A[0],A[1],c='r',s=500)
plt.scatter(B[0],B[1],c='b',s=500)
plt.scatter(c1[0],c1[1],c='k',s=200)
plt.scatter(c2[0],c2[1],c='k',s=200)
plt.scatter(c3[0],c3[1],c='k',s=200)
plt.scatter(c4[0],c4[1],c='k',s=200)
plt.axhline(0)
plt.axvline(0)
plt.axis('scaled')
plt.arrow(A[0],A[1],B[0]-3.1,B[1]-0.3,head_width=0.4,head_length=0.4,fc='k',ec='k',lw=2)
plt.text(A[0]-0.5,A[1]+0.25, "A", fontsize=26)
plt.text(B[0]-0.2,B[1]+0.35, "B", fontsize=26)
plt.text(c1[0]+0.2,c1[1]+0.25, "$D_1$", fontsize=24)
plt.text(c2[0]-0.8,c2[1], "$D_2$", fontsize=24)
plt.text(c3[0]-0.8,c3[1], "$D_3$", fontsize=24)
plt.text(c4[0]+0.1,c4[1]+0.25, "$D_4$", fontsize=24)
plt.xlim(-R-0.2,R+0.2)
plt.ylim(-R-0.2,R+0.2)
plt.xticks(arange(-4.0,4.1,1))
plt.tick_params(axis='both', direction='in', labelsize=18)
plt.tick_params(labeltop=True, labelright=True, labelbottom=True)
plt.tick_params(labelleft=True, bottom=True, top=True, left=True, right=True)
plt.grid(True)
plt.show()


#Necessary functions 
#Calculate brightness at the muon entry point
def entry_brightness(L,c,v,alpha,den):
    tt     = L/c
    aterm  = (c*c*tt*v-L*v*v*cos(alpha))
    bterm  = (v*v*( -L*L*v*v + c*c*L*L + c*c*tt*tt*v*v - 2*c*c*L*tt*v*cos(alpha) + L*L*v*v*cos(alpha)**2))
    xp     = (aterm + sqrt(bterm)) / den
    cterm  = (c*c*v)
    dterm  = (c*c*v*v*v*(tt*v-L*cos(alpha)))
    vp     = (cterm + (dterm/sqrt(bterm))) / den
    kp     = sqrt( L*L + xp*xp - 2*L*xp*cos(alpha) )
    betap  = alpha
    vtp    = vp*sin(betap)
    omegap = vtp / kp
    bp     = abs(omegap/(kp**2))
    return bp

def sec(x):
    return 1/cos(x)
def tan(x):
    return sin(x)/cos(x)

#Different plotting scenarios
def plus_t_vs_x (a, b, color, label):
    plt.plot(a, b, c=color, ls='-' , lw=2.5, label=label)
def minus_t_vs_x(a, b, color, label):
    plt.plot(a, b, c=color, ls='--', lw=2.5, label=label)
def both_t_vs_x (a1, b1, a2, b2, color, label):
    plt.plot(a1, b1, c=color, ls='-' , lw=2.5, label=label)
    plt.plot(a2, b2, c=color, ls='--', lw=2.5)
    plt.xlabel(r'time since muon entry (in ns)',fontsize=18)
    plt.ylabel(r'image distance $x_{pm}$ from entry point (in meters)', fontsize=18)
    plt.axhline(xc, c='k', ls=':')
    
def plus_t_vs_b (a, b, color, label):
    plt.plot(a, b, c=color, ls='-', lw=2.5, label=label)
def minus_t_vs_b(a, b, color, label):
    plt.plot(a, b, c=color, ls='--', lw=2.5, label=label)
def both_t_vs_b (a1, b1, a2, b2, color, label):
    plt.plot(a1, b1, c=color, ls='-' , lw=2.5, label=label)
    plt.plot(a2, b2, c=color, ls='--', lw=2.5)
    plt.axhline(1,c='k',ls=':')
    plt.xlabel(r'time since muon entry (in ns)',fontsize=18)
    plt.ylabel(r'relative brightness ($b/b_{entry}$)', fontsize=18) 
    plt.yscale('log')
    plt.ylim(1e-2,1e+4)
    
def plus_t_vs_ang(a, b, color, label):
    plt.plot(a, b, c=color, ls='-', lw=2.5, label=label)
def minus_t_vs_ang(a, b, color, label):
    plt.plot(a, b, c=color, ls='--', lw=2.5, label=label)
def both_t_vs_ang(a1, b1, a2, b2, color, label):
    plt.plot(a1, b1, c=color, ls='-' , lw=2.5, label=label)
    plt.plot(a2, b2, c=color, ls='--', lw=2.5)
    plt.axhline(phic, c='k', ls=':')
    plt.xlabel(r'time since muon entry (in ns)',fontsize=18)
    plt.ylabel(r'angular locations $\phi_{pm}\;(in\;degrees)$',fontsize=18)
    
def plus_b_vs_ang(a, b, color, label):
    plt.plot(a, b, c=color, ls='-', lw=2.5, label=label)
def minus_b_vs_ang(a, b, color, label):
    plt.plot(a, b, c=color, ls='--', lw=2.5, label=label)
def both_b_vs_ang(a1, b1, a2, b2, color, label):
    plt.plot(a1, b1, c=color, ls='-' , lw=2.5, label=label)
    plt.plot(a2, b2, c=color, ls='--', lw=2.5)
    plt.axvline(phic, c='k', ls=':')
    plt.axhline(1,c='k',ls=':')
    plt.xlabel(r'angular locations $\phi_{pm}\;(in\;degrees)$',fontsize=18)
    plt.ylabel(r'relative brightness ($b/b_{entry}$)', fontsize=18) 
    plt.yscale('log')
    plt.ylim(1e-2,1e+4)


#Computations and plotting
plt.figure()
plotme    = 't vs b' #---------Plot type selection
#plotme    = 't vs x'
#plotme    = 't vs ang'
#plotme    = 'b vs ang'
detector_coordinates = [c1,c2,c3,c4]
colors               = ['k','r','b','g']
detectors            = [1,2,3,4]
labels               = ['Central Detector', "Detector 2",'Detector 3','Detector 4']

for D, color, detector, mylabel in zip(detector_coordinates, colors, detectors, labels):
    
    print("==============================================")
    print("Detector ", detector)   
    AD    = D - A
    BD    = D - B
    L     = LA.norm(AD)
    alpha = arccos((sum(AD*AB))/(L*nAB)) #-Angle between detector and muon track through entry point A
    xc    = L*cos(alpha) - (c*L*sin(alpha))/sqrt(-den) #-Critical height
    print('XC: ',round(xc,3), " m")
    T,XP,XM,BP,BM,PHIP,PHIM = [],[],[],[],[],[],[]
    
    for iii,t in enumerate(times):

        x      = v*t #------------Distance traveled by muon in time t
        ratio  = x/nAB
        X      = array([ (1-ratio)*A[0] + ratio*B[0] , (1-ratio)*A[1] + ratio*B[1] , h   ]) 
        AX,DX  = X - A, X-D
        k      = sqrt(L*L+x*x-2*L*x*cos(alpha)) #--Distance between detector and muon at time t
        t1,t2  = t, k/c
        tt     = t1 + t2 #-----------------Total time taken by detector to see the muon
        
        aterm  = (c*c*tt*v-L*v*v*cos(alpha))
        bterm  = (v*v*( -L*L*v*v + c*c*L*L + c*c*tt*tt*v*v - 2*c*c*L*tt*v*cos(alpha) + L*L*v*v*cos(alpha)**2))
        xp     = (aterm + sqrt(bterm)) / den #------Distance of first cherenkov image from entry point A
        xm     = (aterm - sqrt(bterm)) / den #------------------second----------------------------------
        
        cterm  = (c*c*v)
        dterm  = (c*c*v*v*v*(tt*v-L*cos(alpha)))
        vp     = (cterm + (dterm/sqrt(bterm))) / den #----Velocity of first cherenkov image
        vm     = (cterm - (dterm/sqrt(bterm))) / den #----------------second---------------
        
        kp     = sqrt( L*L + xp*xp - 2*L*xp*cos(alpha) ) #---Distance of first cherenkov image from detector
        km     = sqrt( L*L + xm*xm - 2*L*xm*cos(alpha) ) #---------------second-----------------------------
        
        betap  = pi - arccos( (xp*xp + kp*kp - L*L) / (2*xp*kp) )  
        betam  = pi - arccos( (xm*xm + km*km - L*L) / (2*xm*km) )  
        
        vtp    = vp*sin(betap)   #------Transverse velocity of first cherenkov image
        vtm    = vm *sin(betam ) #-----------------------------second---------------
        omegap = vtp / kp
        omegam = vtm  / km
        bp,bm  = abs(omegap/(kp**2)), abs(omegam/(km **2)) #------Brightness of first and second cherenkov images
        
        phip   = arccos( (L*L + kp*kp - xp*xp ) / (2*L*kp) ) #----Angular location of images as seen by detector
        phim   = arccos( (L*L + km*km - xm*xm ) / (2*L*km) ) #----------------------------do--------------------
        
        XP.append(xp); XM.append(xm); BP.append(bp); BM.append(bm)
        PHIP.append(rad2deg(phip)); PHIM.append(rad2deg(phim)); T.append(tt)

    XP, XM, BP, BM, T, PHIP, PHIM = array(XP), array(XM), array(BP), array(BM), array(T), array(PHIP), array(PHIM)
    TT  = T * 1e+9 #------Convert time in nanoseconds
    conp     = [(XP>=0) & (XP<=nAB)] #---Consider only the images inside the tank
    conm     = [(XM>=0) & (XM<=nAB)] #----------------do-------------------------
    pluslen  = len (XP[conp])
    minuslen = len (XM[conm])
    kc       = sqrt( L*L + xc*xc - 2*L*xc*cos(alpha) ) #---Distance between detector & first RID location
    phic     = rad2deg(arccos( (L*L + kc*kc - xc*xc ) / (2*L*kc) )) #--Angular location of first RID
    if detector==1:
        Bnorm = entry_brightness(L,c,v,alpha,den) #---Use this to normalize brightness wrt the entry point
                                                  #---as seen by the central detector 
    if pluslen == 0:
        if minuslen == 0:
            print("Images outside tank. Skipping...")
        else:
            print('One image moving towards exit B.')
            TT,XM,BM,PHIM = TT[conm], XM[conm], BM[conm]/Bnorm, PHIM[conm]
            if   plotme=='t vs b'  :minus_t_vs_b  (a=TT, b=BM, color=color, label=mylabel)
            elif plotme=='t vs x'  :minus_t_vs_x  (a=TT, b=XM, color=color, label=mylabel)
            elif plotme=='t vs ang':minus_t_vs_ang(a=TT, b=PHIM, color=color, label=mylabel)
            elif plotme=='b vs ang':minus_b_vs_ang(a=PHIM, b=BM, color=color, label=mylabel)
    
    elif pluslen != 0:
        if minuslen == 0:
            print('One image moving towards entry A.')
            TT,XP,BP,PHIP = TT[conp], XP[conp], BP[conp]/Bnorm, PHIP[conp]
            if   plotme=='t vs b'  :plus_t_vs_b  (a=TT, b=BP, color=color, label=mylabel)
            elif plotme=='t vs x'  :plus_t_vs_x  (a=TT, b=XP, color=color, label=mylabel)
            elif plotme=='t vs ang':plus_t_vs_ang(a=TT, b=PHIP, color=color, label=mylabel)
            elif plotme=='b vs ang':plus_b_vs_ang(a=PHIP, b=BP, color=color, label=mylabel)
        else:
            print('Both images moving')
            TTm, XM, BM, PHIM = TT[conm], XM[conm], BM[conm]/Bnorm, PHIM[conm]
            TTp, XP, BP, PHIP = TT[conp], XP[conp], BP[conp]/Bnorm, PHIP[conp]
            if   plotme=='t vs b'  :both_t_vs_b  (a1=TTp, b1=BP, a2=TTm, b2=BM, color=color, label=mylabel)
            elif plotme=='t vs x'  :both_t_vs_x  (a1=TTp, b1=XP, a2=TTm, b2=XM, color=color, label=mylabel)
            elif plotme=='t vs ang':both_t_vs_ang(a1=TTp, b1=PHIP, a2=TTm, b2=PHIM, color=color, label=mylabel)
            elif plotme=='b vs ang':both_b_vs_ang(a1=PHIP, b1=BP, a2=PHIM, b2=BM  , color=color, label=mylabel)

plt.tick_params(axis='both', direction='in', labelsize=18)
plt.legend(prop={'size': 14}) 
plt.show()

\end{lstlisting}
\break

\subsection{C++ Code: Monte Carlo Simulation for fraction of muons showing an RID event}
\lstset{style=mystyle}
\begin{lstlisting}[language=C++]

#include<iostream>
#include<cmath>
#include<cstdlib>
#include<ctime>
using namespace std;

double getRandFloat(double min, double max)
	{
	return min + (double) rand() / (double) (RAND_MAX/(max-min));
	}

int getRandInt(long min, long max)
	{
	return min + ( (long)   rand() / (long)   (RAND_MAX/(max-min+1)) );
	}


double getDistance(double a1[], double a2[], long n)
	{
	double distance=0.0;
	for (unsigned int i=0; i<n; i++)
		distance += pow(a2[i]-a1[i],2.0);
	return pow(distance,0.5);
	}


double getAngle(double b[], double a[], double c[], long n)
    {
    double ab[n], ac[n];
    double origin[3]={0.0,0.0,0.0};
    double dotproduct=0.0;
    for (unsigned int i=0; i<n; i++)
        {
        ab[i] = b[i]-a[i];
        ac[i] = c[i]-a[i];
        }
    for (unsigned int i=0; i<n; i++)
        dotproduct += (ab[i]*ac[i]);
    return  acos (dotproduct/(getDistance(ab,origin,n) * getDistance(ac,origin,n)));
    }


void setCoordinates(long choice, double point[])
    {
    double x,y,z,Y;
    double H=4.5, R=3.65;
    x = -R + ( (double)   rand() / (double)   (RAND_MAX/(R+R)) );
    if (choice==1)
        {
        Y = pow(R*R-x*x,0.5);
        y = -Y + ( (double)   rand() / (double)   (RAND_MAX/(Y+Y)) );
        z = H;
        }
    else if (choice==2)
        {
        z = 0.0 + ( (double)   rand() / (double)   (RAND_MAX/(H)) );
        y = pow(R*R-x*x,0.5);
        int dir = 1 + ( (long)   rand() / (long)   (RAND_MAX/(2-1+1)) );
        if (dir==2)
            y = -y;
        }
    else
        {
        Y = pow(R*R-x*x,0.5);
        y = -Y + ( (double)   rand() / (double)   (RAND_MAX/(Y+Y)) );
        z = 0;
        }
    point[0]=x;
    point[1]=y;
    point[2]=z;
    }


int main(int argc , char* argv[])
    {
    srand(time(0));
    const long   all_cases=atol(argv[1]);
    const long   pi=3.141597;
    const double c=299792458.0/1.33, v=1.33*c, H=4.5, R=3.65;
    const double c1[3] = {0.0, 0.0, 0.0};
    const double c2[3] = {1.85*cos(2*pi/3) , 1.85*sin(2*pi/3), 0.0};
    const double c3[3] = {1.85*cos(4*pi/3) , 1.85*sin(4*pi/3), 0.0};
    const double c4[4] = {1.85*cos(0)      , 1.85*sin(0)     , 0.0};
    long   counts=0, iter=0;
    long   choice_A, choice_B, flag;
    double A[3], B[3], X[3];
    double alpha, nAB, L, xc, ratio;

    while (iter<all_cases)
        {
        choice_A = getRandInt(1,2);
        choice_B = getRandInt(2,3);
        setCoordinates(choice_A, A);
        setCoordinates(choice_B, B);
        while (B[2] > A[2])
            {
            choice_A = getRandInt(1,2);
            choice_B = getRandInt(2,3);
            setCoordinates(choice_A, A);
            setCoordinates(choice_B, B);
            }
        nAB = getDistance(A,B,3);
        double detectors[1][4][3]  = {{{c1[0],c1[1],c1[2]}, {c2[0],c2[1],c2[2]},
                                       {c3[0],c3[1],c3[2]}, {c4[0],c4[1],c4[2]}}};
        flag = 0;
        for (int index=0; index<4; index++)
            {
            alpha = getAngle(B,A,detectors[0][index],3);
            L     = getDistance(A,detectors[0][index],3);
            xc    = L*cos(alpha) - ( (c*L*sin(alpha)) / (pow(v*v-c*c,0.5)) );
            ratio = xc/nAB;
            X[0]  = (1-ratio)*A[0]+ratio*B[0];
            X[1]  = (1-ratio)*A[1]+ratio*B[1];
            X[2]  = (1-ratio)*A[2]+ratio*B[2];

            if ( ((X[0]*X[0]+X[1]*X[1])<=R*R) && (X[2]>=0.0) && (X[2]<=H) )
                {
                counts+=1;
                flag = 1;
                }
            if (flag==1)
                break;
            }
        iter+=1;
        }

    cout<<"Total Particles  : "<<all_cases<<endl;
    cout<<"Particles RIDed  : "<<counts<<endl;
    cout<<"Fraction         : "<<(double)counts/(double)all_cases<<endl;
    return 0;
}
\end{lstlisting}

\end{document}